\documentclass[format=manuscript, review=false, screen=true, 10pt]{acmart}
\fancyhfoffset[LE]{0pt}

\usepackage{booktabs} 

\usepackage[ruled]{algorithm2e} 

\SetAlFnt{\small}
\SetAlCapFnt{\small}
\SetAlCapNameFnt{\small}
\SetAlCapHSkip{0pt}
\IncMargin{-\parindent}

\acmVolume{0}
\acmNumber{0}
\acmArticle{00}
\acmYear{0000}
\acmMonth{0}
\copyrightyear{2018}
\acmArticleSeq{0}

\setcopyright{rightsretained}

\usepackage{comment}
\usepackage[backgroundcolor=yellow]{todonotes}
\usepackage[noend]{algpseudocode}
\usepackage{setspace}
\usepackage{multirow}
\usepackage{xfrac}

\usepackage{amsmath}
\usepackage{mathtools}

\usepackage{subcaption}

\RequirePackage[nomain,nonumberlist,nogroupskip]{glossaries}
\RequirePackage{xspace}
\RequirePackage{bm}
\RequirePackage{amsfonts}
\RequirePackage{textcomp}
\newglossary{symbols}{sym}{sbl}{List of Symbols}

\newglossary{hiddensymbols}{hidsym}{hidsbl}{} 

\newcommand{\sym}[4]{\newglossaryentry{#1}{type=symbols,
        sort=#2,
        name={\ensuremath{#3}\xspace},
        description={#4}}%
    \expandafter\newcommand\expandafter{\csname #1\endcsname}{\ensuremath{#3}\xspace}%
}

\newcommand{\symHidden}[4]{\newglossaryentry{#1}{type=hiddensymbols,
        sort=#2,
        name={\ensuremath{#3}\xspace},
        description={#4}}%
    \expandafter\newcommand\expandafter{\csname #1\endcsname}{\ensuremath{#3}\xspace}%
}

\newcommand{\symLower}[6]{\newglossaryentry{#1#2}{type=symbols,
        sort=#3,
        text={\ensuremath{#4}\xspace},
        name={\ensuremath{#4_{#5}}\xspace},
        description={#6}}%
    \expandafter\newcommand\expandafter{\csname #1\endcsname}[1]{\ensuremath{\ensuremath{#4}\xspace_{##1}}}%
    \expandafter\newcommand\expandafter{\csname #1#2\endcsname}{\ensuremath{\ensuremath{#4}\xspace_{#5}}}%
}

\newcommand{\symLowerComma}[6]{\newglossaryentry{#1#2}{type=symbols,
        sort=#1#2,
        text={\ensuremath{#3}\xspace},
        name={\ensuremath{#3_{#4,#5}}\xspace},
        description={#6}}%
    \expandafter\newcommand\expandafter{\csname #1\endcsname}[1]{\ensuremath{#3}\xspace_{#4,##1}}%
    \expandafter\newcommand\expandafter{\csname #1#2\endcsname}{\ensuremath{#3}\xspace_{#4,#5}}%
}

\newcommand{\symFun}[6]{\newglossaryentry{#1#2}{type=symbols,
        sort=#3,
        text={\ensuremath{#4}\xspace},
        name={\ensuremath{#4\left({#5}\right)}\xspace},
        description={#6}}%
        \expandafter\newcommand\expandafter{\csname #1\endcsname}[1]{\ensuremath{#4}\xspace\left({##1}\right)}%
        \expandafter\newcommand\expandafter{\csname #1#2\endcsname}{\ensuremath{#4}\left({#5}\right)}%
}

\newcommand{\stdname}[1]{{\it #1}}

\sym{Iup}{Iup}{I_{up}}{Mean packet generation interval}
\sym{Len}{Len}{L}{Packet length in bytes}
\sym{tsetup}{ts}{t_s}{Setup time}
\sym{tcooldown}{tc}{t_c}{Cool-down time}
\sym{numpackets}{Np}{N_p}{Number of packets sent per node per run}
\symLowerComma{Nr}{nri}{N}{r}{i}{Number of packets received originating from node $i$}
\symLower{PDR}{i}{PDRi}{\text{PDR}}{i}{Packet delivery ratio for node $i$}

\sym{creq}{creq}{c_{req}}{Required capacity}
\sym{cact}{cact}{c_{act}}{Actual capacity}
\sym{rateout}{mout}{\mu}{Output rate}
\sym{rateoutminloss}{moutminloss}{\mu_{\text{min,a}}}{Minimum output rate according to packet loss rule}
\sym{rateoutmindelay}{moutmindelay}{\mu_{\text{min,b}}}{Minimum output rate according to packet delay rule}
\symLower{p}{t}{pt}{p}{t}{Incoming packets per multi-superframe}
\symLower{ratein}{t}{lt}{\lambda}{t}{Predicted incoming packets per multi-superframe}
\sym{ap}{ap}{\alpha_{\lambda}}{Filter coefficient for packet predictor}
\sym{Tmusu}{Tmusu}{T_{\text{musu}}}{Duration of a multi-superframe}
\sym{maxqueue}{K}{K}{Maximum queue length}
\sym{sq}{s2}{s^2}{Squared coefficient of the variation of the service process}
\sym{PKs}{PKs}{P_{K,\sq}}{Blocking probability}
\sym{pbmax}{pbmax}{p_{b,\text{max}}}{Maximum allowed blocking probability}
\sym{pp}{p}{\rho}{Ratio of input and output rate}
\sym{ppmax}{pmax}{\rho_{max}}{Maximum ratio of input and output rate}
\sym{Wmax}{Wmax}{W_{max}}{Maximum packet delay on a link}

\sym{qsidle}{sidle}{\sigma_{idle}}{Idle State}
\symLower{qs}{qst}{sqst}{\sigma}{q,s,t}{State $q$ packets in the queue, section $s$ and time $t$}

\symLower{qstate}{qi}        {sqi}{\sigma^n}{q,i}{State in the queuing model}

\symLowerComma{Paccept}{Pacceptn}{P}{\text{accept}}{n}{Packet accepting probability}

\symLower{Arr}{n}{An}{\bm{A}}{n}{Random variable for the number of arriving, i.e. generated and received, packets}
\symLower{Que}{n}{Qn}{\bm{Q}}{n}{Random variable for the number packets inserted into the queue}
\symLower{Gen}{l}{Gl}{\bm{G}}{\lambda}{Random variable for the number of generated packets}

\symLower{Delay}{n}         {Dn}{\bm{D}}{n}{Random variable for the queuing delay}

\sym{states}{Sn}{\bm{\Sigma}_{n}}{The set of states for the queuing model}

\sym{lenslot}{Ts}{T_{s}}{Duration of a slot}
\symLower{Rx}{ni}        {bRni}{\beta^{\textsc{RX}}}{n,i}{Probability of receiving a packet in slot $i$}
\symLower{Tx}{ni}        {bTni}{\beta^{\textsc{TX}}}{n,i}{Probability of sending a packet in slot $i$}

\sym{lensched}{lS}{l_{S}}{Overall number of slots in the slotframe}

\sym{Sctn}{S}{\left\vert\mathcal{T}_n\right\vert}{Number of transmission slots in the schedule for node $n$}

\symLower{txslots}{n}        {Tn}{\mathcal{T}}{n}{Transmission slots}
\symLower{rxslots}{n}        {Rn}{\mathcal{R}}{n}{Reception slots}
\sym{unusableslots}{U}{\mathcal{U}}{Unusable slots}

\symLowerComma{txslot}{tni}{t}{n}{i}{Transmission slot}
\symLowerComma{rxslot}{rni}{r}{n}{i}{Reception slot}

\symLower{csteady}{nqi}        {cnqi}{c}{n,q,i}{Probability of being in state $\qstate{q,i}$ in the stationary distribution}
\sym{cvect}{c}{c}{Vector of the $\csteady{n,q,i}$}

\sym{rootnode}{v0}{v_0}{Root node}
\symLower{parent}{n}        {pn}{p}{n}{Parent of node $n$}
\symLower{node}{n}        {vn}{v}{n}{Node with index $n$}

\sym{channels}{c}{\mathcal{C}}{Channels}
\symLower{neighbors}{n}        {Nn}{\text{\bf N}}{n}{Neighbors of node $n$}
\symLower{children}{n}        {Cn}{\text{\bf C}}{n}{Children of node $n$}
\symLower{pdesc}{n}         {gn}{\gamma}{n}{Number of proper descendants of node $n$}
\symLower{counterpart}{n}        {Kn}{\mathcal{K}}{n}{Relation of slots to counterpart}
\symLower{channel}{n}        {Cn}{\mathcal{C}}{n}{Relation of slots to channel}
\symLower{blockedchannels}{n}        {Bn}{\mathcal{B}}{n}{Relation of slots to blocked channels}

\symLower{PER}{bl}{PERbl}{\text{PER}}{b,l}{Packet error rate for a transmission of $b$ bytes}
\sym{Li}		{Li}{{\bf L}_i}{The set of active links during slot $i$}
\symLower{Disturb}{il}{Dil}{{\bf D}}{i,l}{The set of possibly disturbing links during slot $i$}
\symFun{inrange}{vw}{D}{\ensuremath{D}}{v,w}{Predicate that describes potential collisions between nodes}
\symLowerComma{Rup}{n}{\mathfrak{R}}{\text{up}}{v_n}{Reliability that a packet sent by $n$ arrives at the sink}
\symLowerComma{Dup}{n}{\mathfrak{D}}{\text{up}}{v_n}{End-to-end packet delay}

\sym{MO}{MO}{\text{MO}}{Multi-superframe order}
\sym{SD}{SD}{\text{SD}}{Superframe duration}
\sym{MD}{MD}{\text{MD}}{Multi-superframe duration}
\sym{SO}{SO}{\text{SO}}{Superframe order}
\sym{BO}{BO}{\text{BO}}{Beacon order}

\sym{slotCorrection}{tau}{\tau}{Slot correction factor}
\sym{slotsAvailable}{s}{s}{Number of available slots}
\symLower{divisor}{u}        {pu}{\phi}{u}{Divisor for schedule calculation}
\sym{elected}{e}{\epsilon}{Elected neighbor}

\sym{initbexp}      {m0}{m_0}{Initial backoff exponent}
\sym{maxbexp}       {mb}{m_b}{Maximum backoff exponent}
\sym{Sb}            {Sb}{S_b}{Backoff unit time (used as basic time unit)}
\symLower{W}{i} {Wi}{W}{i}{Maximum backoff time span for the $i^{\text{th}}$ channel sensing}
\sym{maxbackoffs}   {m}{m}{Maximum number of channel sensing attempts (\stdname{macMaxCsmaBackoffs})}
\sym{maxretrans}    {n}{n}{Maximum number of retransmission attempts (\stdname{macMaxFrameRetries})}

\newcommand{\mmax}{m_{\text{max}}}
\newcommand{\Pint}{P_{\text{int}}}
\newcommand{\Pintupper}{P_{\text{int,upper}}}
\newcommand{\Pintlower}{P_{\text{int,lower}}}

\newcommand{\ppreimg}{\mathbb{N}_0}

\RequirePackage{colortbl}

\begin{document}

\title[An Analysis of IEEE~802.15.4 DSME]{Reliable Wireless Multi-Hop Networks with Decentralized Slot Management: An Analysis of IEEE~802.15.4 DSME}

\author{Florian Kauer}
\email{florian.kauer@tuhh.de}
\author{Maximilian K\"ostler}
\email{maximilian.koestler@tuhh.de}
\author{Volker Turau}
\email{turau@tuhh.de}
\affiliation{%
  \institution{Hamburg University of Technology}
  \department{Institute of Telematics}
  \streetaddress{Am Schwarzenberg-Campus 3}
  \city{Hamburg}
  \postcode{21073}
  \country{Germany}}

\begin{abstract}
Wireless communication is a key element in the realization of the Industrial Internet of Things for flexible and cost-efficient monitoring and control of industrial processes. Wireless mesh networks using IEEE 802.15.4 have a high potential for executing monitoring and control tasks with low energy consumption and low costs for deployment and maintenance. However, conventional medium access techniques based on carrier sensing cannot provide the required reliability for industrial applications. Therefore, the standard was extended with techniques for time-slotted medium access on multiple channels. In this paper, we present openDSME, a comprehensive implementation of the Deterministic and Synchronous Multi-channel Extension (DSME) and propose a method for traffic-aware and decentralized slot scheduling to enable scalable wireless industrial networks. The performance of DSME and our implementation is demonstrated in the OMNeT++ simulator and on a physically deployed wireless network in the FIT/IoT-LAB. It is shown that in the given scenarios, twice as much traffic can be delivered reliably by using DSME instead of CSMA/CA and that the energy consumption can be reduced significantly. The paper is completed by presenting important trade-offs for parameter selection and by uncovering open issues of the current specification that call for further effort in research and standardization.
\end{abstract}

%
%


%
%

\newlength{\imageh}
\newlength{\imaged}
\newlength{\imagew}

\newcommand{\setimageh}[1]{
 \settoheight{\imageh}{\usebox{#1}}
}

\newcommand{\setimagew}[1]{
 \settowidth{\imagew}{\usebox{#1}}
}

\newcommand{\setimaged}[1]{
 \settodepth{\imaged}{\usebox{#1}}
}

\newcommand{\imd}[1]{
  \newsavebox{\Image}
  \savebox{\Image}{#1}
  \centering\usebox{\Image}\
  \setimageh{\Image}
  \setimagew{\Image}
  \setimaged{\Image}
  \footnotesize
     {\vskip7pt
      The height of the image is : \the\imageh\\
     The width of the  image is : \the\imagew\\
     The depth of the  image is : \the\imaged\\}
}


\maketitle

\renewcommand{\shortauthors}{F. Kauer, M. K\"ostler and V. Turau}

\section{Introduction}

New trends in industrial production, condensed in the term Industry 4.0, call for intelligent solutions for monitoring and control. Conventionally, an industrial plant is composed of one or multiple independent monolithic units. While each unit consists of many tightly coupled sensors, actuators and controllers, only few communication takes place between these units or with superordinate entities.
In the Industrial Internet of Things (IIoT), this approach is revised on two levels. First, large monolithic units are split up in multiple autonomous components. Second, these components are interconnected to exchange information about process control but also machine conditions and performance to optimize the total output. Hence, communication is a core technology for realizing the Industrial Internet of Things \cite{jeschke_industrial_2017}. Especially wireless technologies promise a large flexibility and low costs.
\pagebreak

Such a modular approach allows for fast and cost-efficient adaption to new demands. To demonstrate the significance and the benefit of this approach, we consider a solar tower power plant, where thousands of steerable mirrors reflect sunlight to a central receiver. In existing plants, every motor has a direct wired connection to a single field control unit and is actuated with a high frequency. The authors participated in the research project AutoR where an alternative approach was investigated \cite{pfahl_holistic_2014}. In this proposal, every mirror is a stand-alone unit that can autonomously track the sun. Since this allows for less frequent communication, mainly for power control and state-of-health messages, it is possible to use wireless connections to the mirrors, largely reducing the deployment costs for cabling. Additional mirrors can easily be added without modifications to a wired field bus. For the same reason, using wireless technologies is advantageous when retrofitting an existing plant with new sensors and actuators. This allows for mass customization \cite{kull_mass_2015} that requires accurate monitoring of every workpiece and repeated reconfiguration of the machines. Also, the collection of machine parameters to enable a highly efficient use of the available resources is enabled, for example for predictive maintenance \cite{civerchia_industrial_2017}.

All these applications, however, require a reliable and projectable performance of the wireless communication. Contention based medium access such as CSMA/CA has an inherent potential of packet collisions that is aggravated by an increasing number of nodes and the occurrence of hidden node situations \cite{meiermodel,modelarxiv}. Secondly, the usage of only a single frequency channel wastes potential in terms of throughput and resilience to external interference \cite{gonga_revisiting_2012}.
The emerging extensions to the IEEE~802.15.4 standard, culminating in the publication of the 2015 version of the standard, have high potential to enable a much broader application of energy-constrained wireless mesh networks in the Industrial Internet of Things. The standard introduced techniques for collision-free time slot communication on multiple channels, namely Time Slotted Channel Hopping (TSCH) and the Deterministic and Synchronous Multi-channel Extension (DSME).

Both approaches are able to schedule time and frequency slots to one or multiple communication partners, thus reducing or even avoiding packet collisions. The latter is possible by a sophisticated slot scheduling mechanism that takes hidden node situations into account. While in TSCH, a general outline of the multi-channel slot technique is described that leaves many parts to the upper layers, for example 6TiSCH \cite{ietf-6tisch-architecture-13}, DSME is more specific in terms of slot structure and management. DSME is closely related to guaranteed time slot~(GTS) structure that already existed in earlier versions of IEEE~802.15.4, but extends it to multi-hop and multi-channel networks. It promises collision-free and resilient communication while reducing energy consumption compared to always-on medium access such as CSMA/CA.

While DSME has already found some attention in research as shown in the next section, three aspects call for further attention to enable the widespread usage of DSME in industrial contexts. First, an implementation is required that does not only run in a simulator, but also on actual wireless sensor and actuator networks. Second, the development of a traffic-aware and decentralized slot scheduling to allow for scalable network operation and high performance. And third, a comprehensive analysis of DSME, not only by means of a simulator but also a hardware testbed. This paper tries to fill this gap with the following main contributions:
\begin{itemize}
\item A full-fledged open-source implementation of DSME suitable for simulation and hardware.
\item An assessment of the vital trade-offs for parameterizing a DSME network. 
\item A proposal and analysis for a traffic-aware and decentralized slot management approach.
\item A simulative evaluation showing the performance of DSME in comparison with CSMA/CA.
\item An evaluation of a testbed experiment demonstrating the applicability of DSME on a physically deployed network as well as its reduced power consumption.
\item Analysis of open issues of the current DSME standard and suggestions for improvements.
\end{itemize}

After describing the related work and the basics of DSME in Sect.~\ref{sect:basics}, the influences of the most important DSME parameters are analyzed in Sect.~\ref{sect:parameters}. It is followed by a presentation of our DSME implementation openDSME in Sect.~\ref{sect:openDSME}. Existing scheduling approaches are presented together with the proposed traffic-aware scheduling function used in openDSME in Sect.~\ref{sect:sf}. Evaluations with the OMNeT++ simulator and in the IoT-LAB testbed are presented in Sect.~\ref{sect:simulation} and Sect.~\ref{sect:hardware}, respectively. Open issues of DSME are discussed in Sect.~\ref{sect:issues} and the paper is concluded in Sect.~\ref{sect:conclusion}.

\subsection{Related Work}
In this section, existing research about DSME is presented. Research about slot scheduling as an integral part of a time-slotted medium access is presented in Sect.~\ref{sect:sfrelated}.

An early contribution that analyzes the performance of DSME is \cite{jeong_performance_2012} where DSME with and without CAP reduction is compared to slotted CSMA/CA in a star and square grid configuration. The work determines a higher throughput of DSME than for CSMA/CA while less energy is consumed. This is especially distinct for a large number of nodes. These authors also conducted research on the influence of WLAN interference \cite{lee_performance_2012} and found out that DSME tolerates much higher levels of WLAN interference than CSMA/CA.

Several improvements to DSME were proposed. In \cite{liu_enhanced_2013} an enhanced fast association is proposed that reduces the association time, mainly by spreading the association requests over a longer time period. A new new channel access and beacon broadcast scheme is proposed in \cite{sahoo_novel_2017}, for example by letting nodes only transmit in one CAP period per multi-superframe. Energy enhancements for destination-oriented topologies are proposed in \cite{capone_modeling_2014}. The energy reduction is mainly achieved by turning off the receiver also during CAP phases when no traffic from the parent is expected, indicated by the non-set Frame Pending field in the Enhanced Beacon frame.

The focus of \cite{vallati_improving_2017} is on the analysis of the network setup time, mainly improved by CSMA/CA parameter tuning in the CAP phase and receiving packets also during backoff, a method also proposed for conventional CSMA/CA in \cite{Weigel2015}.
The authors of \cite{hwang_analysis_2014} propose E-DSME for better beacon scheduling by adding permission notifications instead of relying on collision notifications. In \cite{formaldsme} improvements to the slot allocation handshake are proposed after uncovering weaknesses by using a formal analysis triggered by disturbed transmissions during the CAP.

Several studies compare the different enhancements of the IEEE 802.15.4 standard, including \cite{de_guglielmo_ieee_2016} and \cite{juc_energy_2016}. The latter identifies a slight advantage of DSME compared to TSCH in terms of energy consumption. In \cite{alderisi_simulative_2015} DSME and TSCH are compared under scenarios relevant for industrial process automation, showing that DSME is more suitable for larger networks.

While there are several attempts to implement DSME for a simulator, for example for Cooja in \cite{vallati_improving_2017}, for QualNet in \cite{lee_efficient_2016} and for OPNET in \cite{capone_modeling_2014}, there exists, to the best of our knowledge, no publicly available implementation of DSME that can be executed in a simulator as well as on hardware such as wireless sensor nodes. Even more, \cite{de_guglielmo_ieee_2016} explicitly identifies the lack of a complete implementation as the limiting factor for the application of DSME in real environments.

\begin{figure*}[tb]
\centering
	\includegraphics[page=1,width=387.96pt]{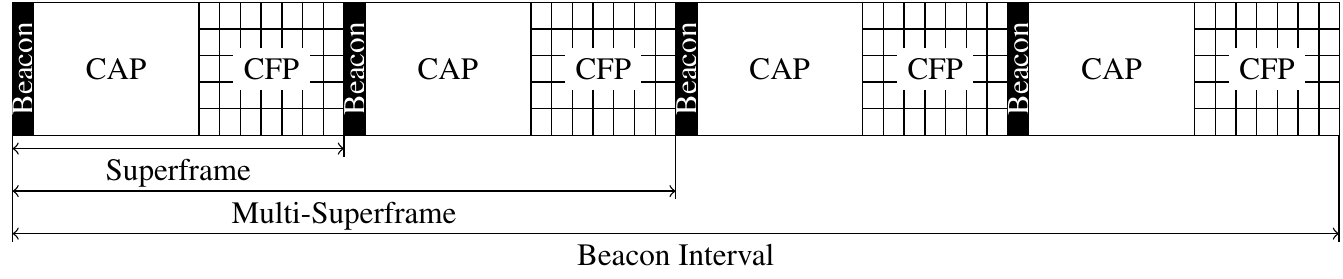}
\caption{DSME Superframe structure for $\SO = 3$, $\MO = 4$ and $\BO = 5$.\label{fig:superframes}}
\end{figure*} %

\section{The Basics of IEEE 802.15.4 DSME}
\label{sect:basics}
This section gives a short introduction into the basic principles of DSME. It describes the predetermined, yet adaptable, slot structure of DSME and the beacon-based approach for time synchronization. This is different from TSCH where the data frames and acknowledgments are used for time synchronization. The main difference, however, is the specification of a bunch of messages and distributed procedures for beacon and slot management in the DSME standard, while in TSCH the management is completely up to the higher layers.

\subsection{Superframe Structure}
\label{sect:structure}
Fig.~\ref{fig:superframes} shows an exemplary superframe structure. Every superframe consists of $\stdname{aNumSuperframeSlots} = 16$ slots of duration
\begin{align}
    \lenslot = \text{\stdname{aBaseSlotDuration}}\cdot2^{\SO},
\end{align}
where \stdname{aBaseSlotDuration} is 60 symbols and $\SO$ is the \stdname{macSuperframeOrder} between 0 and 15, resulting in a superframe duration of
\begin{align}
  \SD = \text{\stdname{aBaseSuperframeDuration}}\cdot2^{\SO},\label{math:SD}
\end{align}
where $\stdname{aBaseSuperframeDuration} = 960 \text{ symbols}$. The first slot of a superframe is a beacon slot (see next subsection), followed by the contention access period (CAP) with 8 slots and a contention free period (CFP) with 7 time slots for allocating guaranteed time slots (GTS). During the CAP, contention based channel access is used. The beacon and the messages in the CAP are sent on a predefined, common channel.

A GTS is dedicated for communication between two nodes on a given channel, even simultaneously to another GTS if the links are spatially separated or a different channel is used. Every GTS is repeated with an interval of one multi-superframe. A multi-superframe consists of $2^{\MO-\SO}$ superframes with the multi-superframe order $\MO$ between $0$ and $22$. For getting more GTS per time, a CAP reduction mode is defined where only the first superframe of a multi-superframe has a CAP and all other superframes have an extended CFP instead. Finally, the beacon interval defines the repetition interval of a beacon. With the beacon order $\BO$ the number of multi-superframes in a beacon interval is $2^{\BO-\MO}$.

In contrast to TSCH, where slot frames of arbitrary length can be constructed, the number of slots per multi-superframe is restricted to certain values given by
\begin{align}
    7\cdot 2^{\MO-\SO} \label{eq:gtsnocapred}
\end{align}
or when using the CAP reduction mode
\begin{align}
    7 + 15\cdot \left(2^{\MO-\SO}-1\right) \label{eq:gtscapred}
\end{align}
GTS are available depending on the integer $\MO$ and $\SO$.

\subsection{Time Synchronization}
\label{sect:timesync}
The time synchronization relies on beacons just like in IEEE 802.15.4 star networks. What is different is that multiple nodes can send beacons. This is necessary for multi-hop networks to distribute a common notion of time throughout the network. Before joining the network, a node scans for beacons and associates via the sender of a beacon. After that, it tracks the beacon of its time synchronization parent, that is not necessarily equal to the parent in a routing tree, to compensate its clock drift. It is important to not listen to beacons of other devices, because this can easily lead to cyclic groups of nodes that drift away from the global notion of time.

A node can reserve a beacon slot to send an own beacon within its neighborhood by selecting a slot and sending a beacon allocation notification command as broadcast. A receiving node will either mark this beacon slot as occupied and will include this information in the enhanced beacons sent out by it or, if a conflict is detected, it sends a beacon collision notification to reject the selection of this beacon slot.

\subsection{Slot Allocation Handshake}
\label{sect:gts}
\begin{figure}[h]
\centering
    \includegraphics[width=162.64pt]{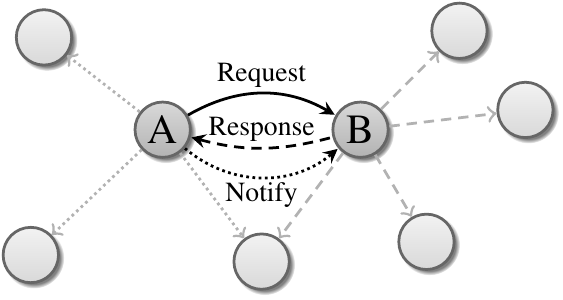}
\caption{GTS allocation handshake.\label{fig:allocation}}
\end{figure}

For scheduling beacons in a distributed way, the IEEE 802.15.4 standard specifies a message exchange consisting of three steps as depicted in Fig.~\ref{fig:allocation}.

\begin{enumerate}
  \item For reserving one or multiple slots for usage on a link between nodes A and B, node A sends out a \textbf{DSME GTS Request} to B. It includes a slot allocation bitmap (SAB) that indicates the node's available time slots and channels where all slot-channel combinations used by neighbors are marked as unavailable as well as all channels for the time slots used by A, because A can not serve two simultaneous slots on different channels. The request also includes a preferred superframe and a slot ID.
  \item After node B received the request, it merges the received SAB with the local information about available slots and selects an available GTS, possibly respecting the preferred slot. This selection is sent back to node A as \textbf{DSME GTS Response} (formerly \emph{Reply} in IEEE 802.15.4e). By using a link-layer broadcast, all neighbors of B will be informed about the selection, depicted by the dashed arrows. They store this information for their next allocations.
  \item Finally, the response is acknowledged by sending a \textbf{DSME GTS Notify}. This is also sent as broadcast, indicated by the dotted arrows, because A and B do not necessarily share the same neighbors.
\end{enumerate}

If a slot collision occurs, especially when a response or notify was not correctly received by a node that afterwards initiates an own allocation procedure for the same slot, a \textbf{DSME GTS Request} with management type \textbf{Duplicate allocation notification} is sent to invalidate the handshake.  
The GTS handshake is also applied when a slot shall be deallocated, for example if the scheduler requests a reduction in the number of slots or that no successful communication took place in this slot for \stdname{macDsmeGtsExpirationTime} multi-superframes in a row. This implies that either no packets were scheduled to be sent or the transmission was not successful, for example due to external interference.

The decision on how many slots are to be requested and when to do this is handled by a higher layer. A new approach that takes the current amount of traffic into account is presented in Sect.~\ref{sect:sfimpl}.
\pagebreak

\section{Parameterization}
\label{sect:parameters}

Several trade-offs have to be made to adapt DSME to a given application. The most important ones are presented in this section.

\vspace{-0.2cm}

\subsection{Superframe Order and Multi-Superframe Order}

As described in Sect.~\ref{sect:basics}, the $\SO$ directly influences the slot length. To transmit full-length IEEE~802.15.4 packets of $127\,\text{Byte}$, $\SO$ must be at least $3$. If only shorter messages are transmitted, a smaller $\SO$ can increase the throughput in terms of packets per time. A larger $\SO$ can on the other hand be used to decrease the power consumption for networks with low traffic by turning off the transceiver after sending or receiving a packet. Another option is to send multiple packets per slot, but that is out of the focus of this work, as we expect the gain to be low and the number of upcoming problems to be high.

The multi-superframe order $\MO$ is mainly responsible for setting the number of distinct slots according to the calculations at the end of Sect.~\ref{sect:structure}. Since the multi-superframe is repeated in time, this number determines the granularity of the slot allocation. For example, with one superframe per multi-superframe, only 7 distinct slots can be assigned. No more than 7 nodes per neighborhood can allocate slots and they can only coarsely adapt the number of allocated slots to the traffic. With more slots, more nodes can be handled and the granularity is finer, but more management traffic is required to transmit the same number of packets per time.

Furthermore, the multi-superframe order $\MO$ determines the overall share of the CAP if CAP reduction is applied. With more superframes per multi-superframe and enabled CAP reduction, less time is available for management traffic, but more time is available for transmitting payload data via guaranteed time slots.

\vspace{-0.2cm}

\subsection{Communication in the CAP}
In openDSME, unslotted CSMA/CA is used within the CAP, though slotted CSMA/CA would be possible since beacons are available. However, the complexity of slotted CSMA/CA is higher and it usually does not provide better performance, as shown for example in \cite{wang_analysis_2009}. Special considerations have to be made for DSME due to the CFP and the beacon slots. For example when issuing a transmission towards the end of the CAP, it might not be possible to complete the backoff, clear channel assessment (CCA), data transmission and potentially ACK transmission within the remainder of a CAP. Therefore, the CCA is postponed to the next CAP. In fact, the backoff is increased by the duration of the intermediate CFP and the beacon slot to avoid that all CCAs take place directly at the beginning of the CAP. This will be repeated if the backoff was selected to be even longer than a single CAP.

Another consideration in this context is the selection of the \stdname{macMinBE} $\initbexp$ that determines the interval of $\left[0,\left(2^{\initbexp}-1\right)\cdot\Sb\right]$ symbols from which the first backoff is randomly selected, where
\begin{align*}
\Sb\coloneqq\text{\stdname{aUnitBackoffPeriod}}\cdot\text{Symbol duration}\!=\!20\cdot 16\,\text{\textmu s}.
\end{align*}
A large share of the messages sent in the CAP are DSME GTS requests issued by the slot scheduler running before the CAP, hence many backoffs start at the beginning of the CAP. This especially holds for all messages not generated during the CAP. If the backoffs are too short, the probability of collisions is very high, especially in hidden node situations. In Table~\ref{tbl:capbackoff}, the length of the CAP is compared to the length of the maximum initial backoff for various settings. A value of $\initbexp$ of at least $5$ should be chosen to spread out the beginning of the transmissions in the first $16\%$ of the CAP for $\SO=3$ in order to mitigate this problem.

\begin{table}
\footnotesize
\centering\begin{tabular}{ | c | r | r | r | r | }
    \hline
    \multirow{ 2}{*}{\textbf{\SO}}& \multicolumn{2}{c|}{\textbf{Slot Duration}} & \multicolumn{2}{c|}{\textbf{CAP Duration}} \\\cline{2-5}
     & \multicolumn{1}{c|}{\textbf{Symbols}} & \multicolumn{1}{c|}{\textbf{ms}} & \multicolumn{1}{c|}{\textbf{Symbols}} & \multicolumn{1}{c|}{\textbf{ms}}\\\hline
     1 & 120 & 1.92 & 960 & 15.36 \\\hline
     2 & 240 & 3.84 & 1920 & 30.72 \\\hline
     3 & 480 & 7.68 & \!\!\!\cellcolor{lightgray}3840 & 61.44 \\\hline
     4 & 960 & 15.36 & 7680 & 122.88 \\\hline
     5 & 1920 & 30.72 & 15360 & 245.76 \\\hline
     6 & 3840 & 61.44 & 30720 & 491.52 \\\hline
\end{tabular}
\hspace{0.2cm}
\begin{tabular}{ | c | r | r | }
    \hline
    \multirow{ 2}{*}{\boldmath$\initbexp$} & \multicolumn{2}{c|}{\textbf{Max. Initial Backoff}}\\\cline{2-3}
  & \multicolumn{1}{c|}{\textbf{Symbols}} & \multicolumn{1}{c|}{\textbf{ms}}\\\hline
            3 & 140 & 2.24 \\\hline
            4 & 300 & 4.80 \\\hline
            5 & \!\!\!\cellcolor{lightgray}620 & 9.92 \\\hline
            6 & 1260 & 20.16 \\\hline
            7 & 2540 & 40.64\\\hline
            8 & \!\!\!\cellcolor{lightgray}5100 & 81.60 \\\hline
\end{tabular}
  \vspace{0.4cm}
    \caption{Comparing the slot and CAP duration to the maximum length of the initial backoff phase for unslotted CSMA/CA. For example, for $SO=3$ and $\initbexp=5$, the CCAs of transmissions issued at the beginning of the CAP will take place in the first $16\%$ of the CAP, while for $\initbexp=8$, this phase will even continue in the subsequent CAP.\label{tbl:capbackoff}}
  \vspace{-0.6cm}
\end{table}

Finally, the CSMA/CA settings influence the selection of the \stdname{macResponseWaitTime} that is the maximum time until a response, for example a DSME GTS Response or Notify, is considered as lost. The reasons for this are further elaborated in \cite{formaldsme}. It is given in multiples of 960 symbols (\stdname{aBaseSuperframeDuration}). The maximum time in symbols it can take until a packet of $S$ symbols ($\sfrac{S}{2}\,\text{Bytes}$) is successfully transmitted is calculated as
\begin{align*}
\maxretrans \cdot \left(\left(\sum_{i=0}^{\maxbackoffs} 20\cdot\W{i}\right) + \stdname{aCcaTime} + S + \stdname{macAckWaitDuration}\right),
\end{align*}
where $\maxretrans$ is \stdname{macMaxFrameRetries}, $\maxbackoffs$ is \stdname{macMaxCsmaBackoffs}, \stdname{aCcaTime} $= 8$, \stdname{macAckWaitDuration} $= 54$ as defined in the IEEE~802.15.4 standard and
\begin{align}
\W{i} = \begin{cases}
2^{\initbexp} & i = 0 \\ 
2^i\cdot \W{0} & 0 < i \leq \maxbexp-\initbexp\\
2^{\maxbexp-\initbexp}\cdot \W{0} & i > \maxbexp-\initbexp
\end{cases},
\end{align}
where $\maxbexp$ is \stdname{macMaxBE}. With $\maxretrans=3$, $\maxbackoffs=4$, $\initbexp=5$, $\maxbexp=7$ and $S=50$, this results in $29106$ symbols, that is about $7.6$ CAP phases for $\SO=3$, so without CAP reduction it takes up to $931\,\text{ms}$ until the message is delivered, resulting in a $\stdname{macResponseWaitTime}$ of $61$. For CAP reduction, this value is multiplied by the number of superframes per multi-superframe, so we use $244$ for the evaluation. Of course, it rarely happens that the full time is taken, but the default value of $32$ and even the maximum of $64$ given by the standard are too low for many practical applications, especially when using large $\SO$ and $\MO$ and since this calculation does not even consider processing delay for parsing the request and assembling the response.

\subsection{macDsmeGtsExpirationTime}
As mentioned in Sect.~\ref{sect:gts}, slots expire either if they are not used or if many transmission errors occur. The destination device of a GTS will count the number of GTS occurrences with no reception and start a deallocation when the counter reaches \stdname{macDsmeGtsExpirationTime}. Furthermore, the source device counts the number of missed acknowledgments in a row and uses the same threshold to issue a deallocation. Since every GTS occurs once per multi-superframe, packets have to be transmitted to avoid deallocation with an interval length of at most
\begin{align}
  \stdname{macDsmeGtsExpirationTime} \cdot 2^{\MO-\SO} \cdot \SD.
\end{align}

If the packet sending interval is too large or has a high variance, repeated deallocations and allocations will induce a high overhead. Thus, the default value of $7$ is often too small. For example with $\MO=4$, the GTS expires after $1.72\,\text{s}$. For our experiments we significantly increase this value to $50$ so reallocations would only take place when no packets are sent within an interval of $12.29\,\text{s}$ for $\MO=4$.

A very high value for \stdname{macDsmeGtsExpirationTime}, however, comes with the disadvantage that invalid GTSs will be recognized later and it takes longer until idle GTSs will be deallocated after changes in the topology. The first is mitigated by using the method proposed in \cite{formaldsme}, but external interferences (e.g. WiFi) during the CFP will still affect the performance in this case.

\subsection{Synchronization and Coordinator Assignment}
The beacon order $\BO$ is important for the synchronization and association in a multi-hop network. Since every beacon is repeated once within the beacon interval, a high $\BO$ will lead to a longer time until a node receives a beacon. Traditionally, according to the IEEE~802.15.4 standard, scanning has to be done on all available channels, so scanning will take very long to collect all beacons. Even after the association, a very large beacon interval can lead to problems if the clock drift of the nodes is too large. On the other hand, the $\BO$ determines the number of distinct beacon slots, so with a small $\BO$ there might not be enough beacon slots available for all nodes to be coordinator, even considering the possibility of spatial reuse.

In order to improve the scanning and association phase, new techniques were proposed \cite{liu_enhanced_2013,nam_enhanced_2014}, including the reduction to a predefined set of channels to scan. In order to allow for a shorter beacon interval in dense networks, it is also possible to let only a subset of the nodes become coordinator. The procedure implemented in openDSME will be presented in Sect.~\ref{sect:formation}.

\section{openDSME}
\label{sect:openDSME}
This section presents our implementation of the IEEE 802.15.4 DSME, called openDSME. 
While it was motivated by the application of DSME in an actual industrial plant in the context of the AutoR research project \cite{pfahl_holistic_2014}, it also serves as a comprehensive tool for conducting research in the context of DSME.
Therefore, openDSME can be executed on wireless hardware, for application in real-world scenarios, as well as in simulators, allowing for execution in reproducible and controlled conditions. Furthermore, using the same source code for hardware and simulation is especially useful for convenient development and debugging.

It is also not restricted to a specific framework and is designed with portability in mind, so it can be plugged into many different systems that provide an interface to simulated or real radio hardware and upper network and application layers.
This is different than most existing implementations for example of TSCH that are either tightly coupled with an existing framework as the TSCH implementation in Contiki or are built from scratch including the required upper and lower layers as openWSN \cite{watteyne_openwsn:_2012}. There already exist integrations of openDSME into the OMNeT++ simulator \cite{varga2001omnet++}, CometOS \cite{unterschutz2012cross} including support for the ATmega256RFR2 and Contiki \cite{dunkels_contiki_2004} including the Cooja simulator and the M3OpenNodes provided by the FIT IoT-LAB~\cite{iotlab}.

The current implementation of openDSME, published as open-source software\footnote{\url{http://opendsme.org}}, consists of about 18000 lines of C++ source code (including comments), compared to about 2000 lines of code for the CSMA/CA implementation of the INET framework \cite{INET} used for comparison in the simulative evaluation. The higher amount of code for openDSME comes from the higher complexity required for a distributed time-slotted medium access compared to one based on carrier sensing.

\subsection{Software Structure}
\begin{figure}[htb]
\centering
        \includegraphics[width=179.14pt]{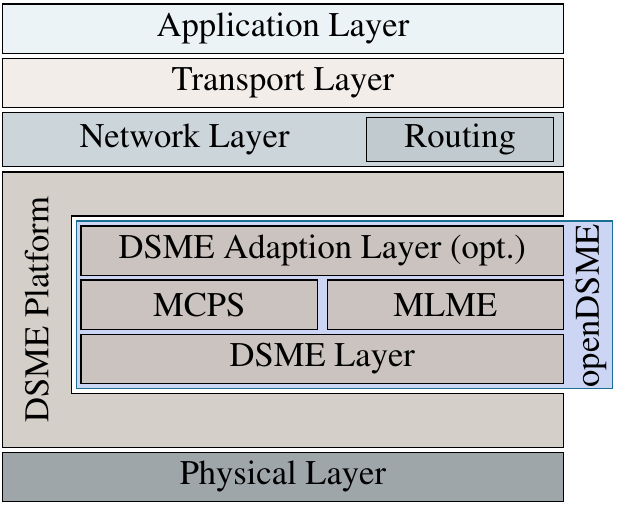}%
\caption{The communication stack with the embedded module structure of openDSME.\label{fig:openDSME}}
\end{figure}

Fig.~\ref{fig:openDSME} shows the basic structure of openDSME and how it is embedded in a network stack. The main part of DSME is implemented in the DSME layer. It sends and receives its messages from the lower layers and communicates with the upper layers via the MCPS and the MLME interfaces. As stated above, DSME leaves some tasks for upper layers, most prominently the decision about how many slots should be allocated. All these features that are required for the full functionality are integrated in openDSME, but separated from the actual DSME implementation as optional DSME Adaption Layer. The latter includes, for example, the GTS Helper that is, together with the interchangeable scheduling module, responsible for selecting the GTS slots and requesting a GTS handshake. It is then performed by the GTS Manager in the DSME Layer. By this approach it is possible to use either pure DSME that is controlled by the standardized MCPS and MLME interfaces or a simple packet interface that hides most complexity. While the first approach supports flexible adaptation to the application, the second approach provides a drop-in replacement for existing link-layer implementations.

A thin wrapper, called DSME Platform, adapts the interface of openDSME to the respective platform. Most prominently it wires the message interfaces as well as it provides a basic timer for openDSME. It connects openDSME to upper layers such as the network and application layer as well as to the physical layer interface of the actual hardware or the simulator. Only basic functionality is required by the physical layer, including packet transmission and reception and carrier sensing, but functionality such as backoffs and acknowledgments is implemented in software, in particular the ACK and the CAP layer. On the upside, this allows for a very flexible implementation. Especially under consideration of inefficient hardware MAC layers \cite{Weigel2015}, this is a large bonus and furthermore allows for easy adaptation to other platforms. On the downside, for hardware with little computational power, timing issues become relevant since DSME poses real time requirements for delays such as the maximum wait time for an ACK, while for simulations this is mostly irrelevant since the simulation time is usually decoupled from the real-world time.

\section{Slot Scheduling}
\label{sect:sf}
While DSME defines mechanisms to allocate and deallocate slots in a distributed fashion as described in Sect.~\ref{sect:gts}, it does not specify to whom and how many slots should be allocated.
This is the task of an upper layer that has to take the requirements of the IIoT application into account.
First, a sophisticated schedule is required for a network with a high traffic demand to enable an efficient usage of the available resources, while simple schedules are sufficient for networks with low throughput. Secondly, traffic with high fluctuation should be accounted for on-demand, while for constant traffic a fixed prereservation is usually sufficient. Third, also the volatility of the topology is important for the required reactivity of the scheduling approach. For example, in networks with many nodes that cover a large area, frequent changes to the schedule are unavoidable. This is even more relevant in case of mobility.

In the Industrial Internet of Things, all combinations of requirements can be found. While we expect that the presented approach can cope with a lot of diverse requirements, it is optimized for scenarios where a lot of traffic that follows a stationary random distribution has to be delivered reliably.
In the simplest case this is a fixed transmission interval that is common for tracking machine parameters such as the state-of-health. One example is the previously mentioned solar tower power plant where a constant monitoring of the steerable mirrors is required in order to maintain the power output of the plant. In other applications, random distributions such as a Poisson distribution might occur. Random processes are for example common for mass customization applications where workpieces on conveyor belts are monitored. 
Finally, in the presented approach, topology changes such as fluctuating wireless conditions will be handled by a decentral recalculation of the schedule, but high mobility is problematic where slot allocations are only valid for a short time-span.
The next section presents existing approaches, followed by a description and analysis of the method implemented in openDSME.

\vspace{-0.2cm}
\subsection{Scheduling Techniques for Multi-Hop TDMA}
\label{sect:sfrelated}
Methods for slot scheduling are a broadly studied topic. The existing approaches can be coarsely categorized in centralized and decentralized approaches. In a centralized approach such as \cite{pottner_constructing_2014,palattella_optimal_2013} statistics about every link in the network are collected at a common entity. It then calculates an adequate schedule to be distributed in the network. With the global knowledge it is easier to calculate optimized schedules. However, the collection of the network state at a single entity comes with a high overhead and is especially unsuitable for very large networks. This is even worse if the topology or the link quality changes because a repetition of the whole process is usually required.

In decentralized approaches such as \cite{tinka_decentralized_2011,accettura_decentralized_2015} the schedule is constructed on the basis of local decisions. This may not lead to globally optimal schedules, but theses approaches are especially useful for large or volatile networks.
There are also hybrid approaches such as \cite{modeltdma,hwang_distributed_2017} where a distributed algorithm is used that does not require a central entity to calculate the schedule, but the global network topology is considered nevertheless. This way the overhead is reduced, but a single broken link might still influence the schedule in large parts of the network.

Another distinction between the approaches is the consideration of heterogeneous traffic. While approaches without traffic-awareness can often be realized quite easily and even without any management traffic, such as Orchestra \cite{orchestra}, the achievable throughput is usually much lower than for traffic-aware schedules as shown in \cite{modeltdma}. This negative effect is also known as funneling effect \cite{wan_siphon:_2005} in networks with only one (or few) sinks. Thus, it is advised to take for example the estimated traffic over a link \cite{palattella_--fly_2016} or the current queue level \cite{domingo-prieto_distributed_2016} into account.

Other promising features for scheduling include traffic isolation \cite{theoleyre_experimental_2016} or the minimization of the energy consumption \cite{ojo_energy_2017}.
In \cite{lee_distributed_2017}, important properties of existing methods are analyzed and compared. The selection of the method used in an actual application highly dependents on the requirements. There is, however, an ongoing effort to agree on standardized approaches to improve interoperability and reusability, leading for example to SF1 \cite{sf1} and SFX \cite{sfx}, the latter being a standardization draft of the approach presented in \cite{palattella_--fly_2016}.

\vspace{-0.2cm}
\subsection{Traffic-Aware Decentralized Slot Management}
\label{sect:sfimpl}
In the following, a method for traffic-aware decentralized slot management is proposed. It requires no explicit information from the routing layer, in contrast to \cite{hwang_distributed_2017}, and it also does not require any additional message exchange apart from the slot allocation handshake of DSME. This makes it both scalable and versatile so it can be used with any routing layer, including RPL \cite{rfc6550} and GPSR \cite{karp_gpsr:_2000} as shown in the evaluation.
Nevertheless, it can dynamically adapt to the current traffic load by predicting the future amount of traffic per link. It shares some similarities with \cite{palattella_--fly_2016}, especially the utilization of overprovisioning, but in contrast, the approach presented in the following is not prone to slot collisions due to the slot allocation handshake. Furthermore, we shed more light on the characteristics of the traffic predictor.

In the following, every outgoing link of a node is considered separately. The number of packets pushed to the sending queue during one multi-superframe $\pt$ is counted. This includes the traffic generated at that node and the traffic to be forwarded. An exponentially weighted moving average filter is applied to this value to predict the traffic load $\rateint$ in future multi-superframe according to
\begin{align}
\rateint = \ap \cdot \pt + (1-\ap) \cdot \ratein{t-1}, \text{ with } \ratein{0} = 0.\label{eq:smooth}
\end{align}
This corresponds to the average number of transmission time slots per multi-superframe for the considered neighbor. Accordingly, DSME GTS Requests can be sent to either allocate or deallocate slots to match the actual number of slots $\cact$ with the required number of slots $\creq$. However, directly setting $\creq = \left\lceil\rateint\right\rceil$ would lead to many repeated slot allocations and deallocations (see Fig.~\ref{fig:tps}). Therefore, a hysteresis is applied, so the required number of slots is calculated as
\begin{align}
    \creq = \begin{cases}
        \left\lceil\rateint\right\rceil & \text{for } \rateint - \cact > 0\\
        \left\lceil\rateint\right\rceil + 1 & \text{for } \rateint - \cact < -2\\
        \cact & \text{else.}\\
    \end{cases}
\end{align}

\begin{figure}[htb]
\centering
        \includegraphics[width=339.62pt]{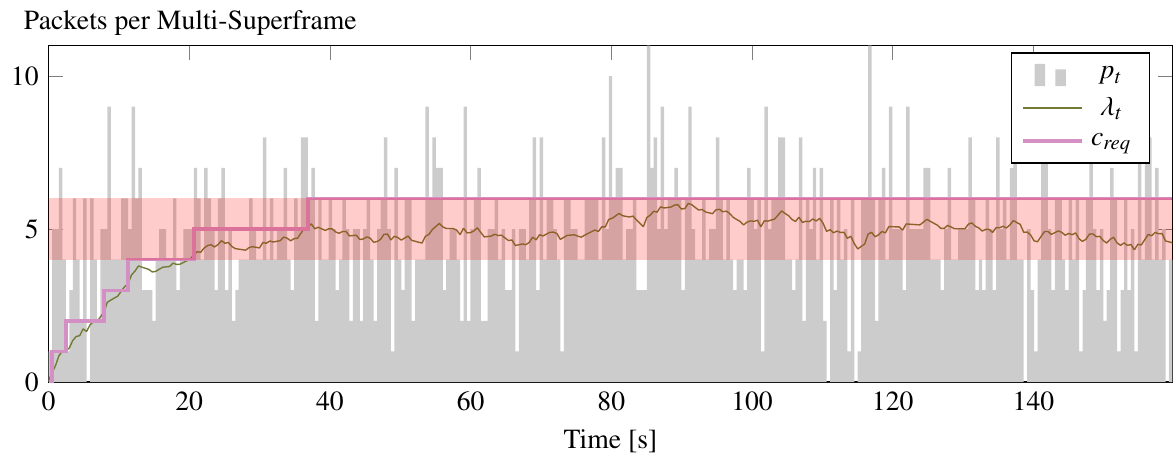}%
\caption{An exemplary course of the presented traffic-aware and decentralized slot management.\label{fig:controllercourse}}
\end{figure}

An exemplary course is depicted in Fig.~\ref{fig:controllercourse}. The gray bars exemplarily represent Poisson distributed traffic with mean $\mu=5$ packets per multi-superframe with $\SO=3$ and $\MO=5$. $\rateint$ initially increases and then fluctuates around $\mu$. For the plot, $\ap = 0.05$ is chosen, so $\rateint$ stays, at least in the given time section, within the hysteresis interval of $\mu\pm1$ depicted by the horizontal bar after about $20\,\text{s}$. A stable configuration is reached after $40\,\text{s}$.

\subsection{Influence of $\ap$}
\label{sect:alpha}
The parameter $\ap$ is used to control the amount of smoothing. For a $\ap$ close to $1$, the current value has a large influence. The system reacts faster to changes in the amount of traffic, but $\rateint$ has a high fluctuation, leading to many GTS allocations and deallocations. For a $\ap$ close to $0$, the opposite holds. Due to the hysteresis, no GTS allocations or deallocations take place if $\rateint$ stays within the respective interval.
This effect is assessed in the following for a generic traffic distribution. The number of incoming packets per multi-superframe is described by a sequence of independent random variables $X_j$ that follow a probability distribution with mean $\mu$ described by the probability mass function
\begin{align}
f_X&: \ppreimg \to \left\{x \in \mathbb{R} \,\middle|\, 0 \leq x \leq 1\right\}.
\end{align}

The random variable $Y_n$ describes the value of $\rateint$ after $n$ steps, starting with $0$ and continuing with
\begin{align}
Y_n &= \ap \cdot X_j + \left(1-\ap\right) \cdot Y_{n-1}.
\end{align}

The probability mass function of $Y_n$ is given by
\begin{align}
    f_{Y_n}&: K_n \to \left\{x \in \mathbb{R} \,\middle|\, 0 \leq x \leq 1\right\}, \text{with}\\
    K_{n} &= \left\{ \ap\cdot k + \left(1-\ap\right)\cdot m \,\middle|\, k \in K_{n-1}, m \in \ppreimg \right\}\\
    K_0 &= {0}, f_{Y_0}(0) = 1\\
    f_{Y_n}(k) &= \sum_{i\,\in\,\ppreimg} f_{X}(i) \cdot f_{Y_{n-1}}\left(\frac{k-(1-\ap)\cdot i}{\ap}\right).
\end{align}

The probability mass function after the settling time is given by $f_{Y_\infty} = \lim_{n \to \infty} f_{Y_n}$ and $\Pint$ is the probability that $\rateint$ is within the interval $\left[\mu-1,\mu+1\right]$ at a given point in time after the initial settling is given by
\begin{align}
\Pint = \sum_{k\,\in\,K_{\infty}} \begin{cases}
    f_{Y_\infty}(k) & \mu-1 \leq k \leq \mu + 1\\
          0 & \text{else.}
      \end{cases}
\end{align}

However, the actual evaluation is computationally infeasible, so an algorithm to calculate an upper and a lower boundary for $\Pint$ is presented in Appendix \ref{sect:ewmacalculations}. The second effect to consider is the delay induced by the smoothing. Given a fixed packet rate $\mu$, the number of multi-superframes it takes until $\mu-1$ is reached is given by%
\begin{align}
t &= \frac{\log\left(\mu^{-1}\right)}{\log\left(1-\ap\right)}
\end{align}
as derived in Appendix \ref{sect:ewmacalculations}.

\begin{figure}[bth]
\centering
        \includegraphics[width=192.82pt]{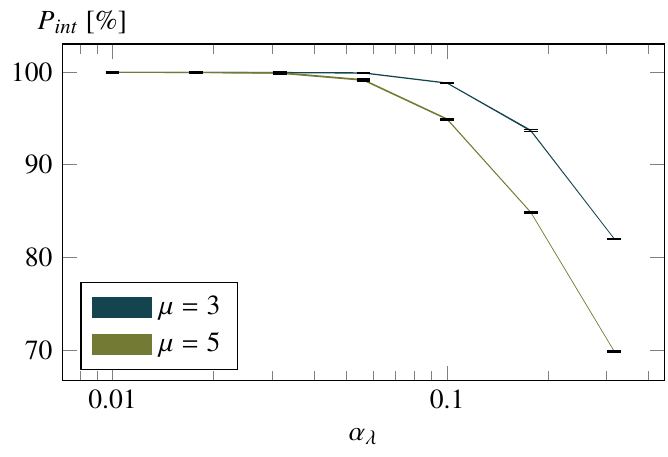}%
    \hspace{0.1cm}
    \includegraphics[width=188.82pt]{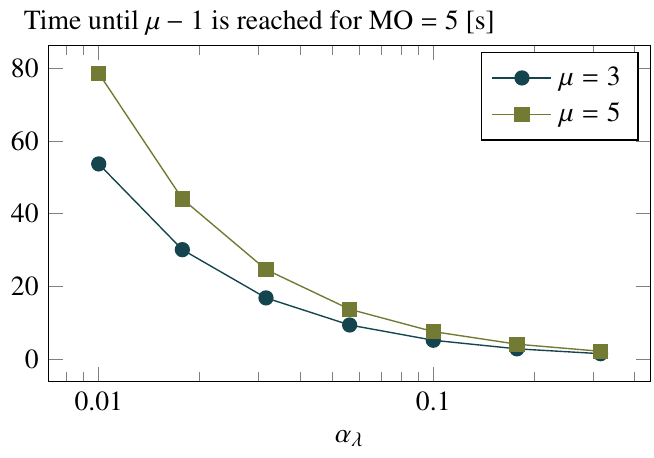}
\caption{Trade-off for choosing the $\ap$ parameter.\label{fig:alphatradeoff}}
    \vspace{-0.2cm}
\end{figure}

For a fixed transmission interval, the calculation always yields $\Pint=100\%$. This is obvious since $\rateint$ stays constant after the initial settling. For a Poisson traffic distribution, the results are more interesting as shown in Fig.~\ref
{fig:alphatradeoff}. With $\ap \leq 0.05$, the probability of staying within the hysteresis interval is larger than $99\%$, but decreases for larger $\ap$.
In the same range, it takes more than $10\,\text{s}$ to react to significant traffic increases, either at the initialization or later, for example when routes change. In the following, $\ap = 0.05$ is applied, but larger $\ap$ should be chosen for a more reactive system or a smaller $\ap$ to reduce the management traffic.

\subsection{Depreciate Links}
\label{sect:depre}
While the presented approach leads to a stable slot assignment under static conditions and can still adapt to changes in the traffic load or the radio environment, its hysteresis leads to an unwanted effect if the connection between two nodes is lost. This might be due to changing channel conditions or mobility. If the network layer detects that a link does no longer exist, no more traffic will be routed over this link ($\pt = 0$). According to Eq.~(\ref{eq:smooth}), however, $\rateint > 0$ will always hold and thus $\creq$ will be at least $1$ even if the link is no longer available. Therefore, the node will repeatedly try to allocate at least a slot for this link. Is is especially unwanted if the link is too bad to allow for a successful slot allocation handshake that will therefore be retried again and again.

To avoid this effect, a counter is introduced that counts the number of multi-superframes since the last multi-superframe with packets to be sent out ($\pt > 0$). If this counter reaches a threshold, $\creq$ is forced to $0$. The threshold depends on the traffic distribution and the stability of the network, but if the amount of traffic per time is not too low, setting it to the same value as \stdname{macDsmeGtsExpirationTime} is usually an good choice. The normal operation is resumed as soon as packets are available again.

\begin{figure}[b]
\begin{minipage}{0.48\textwidth}
\centering
        \includegraphics[width=183.6pt]{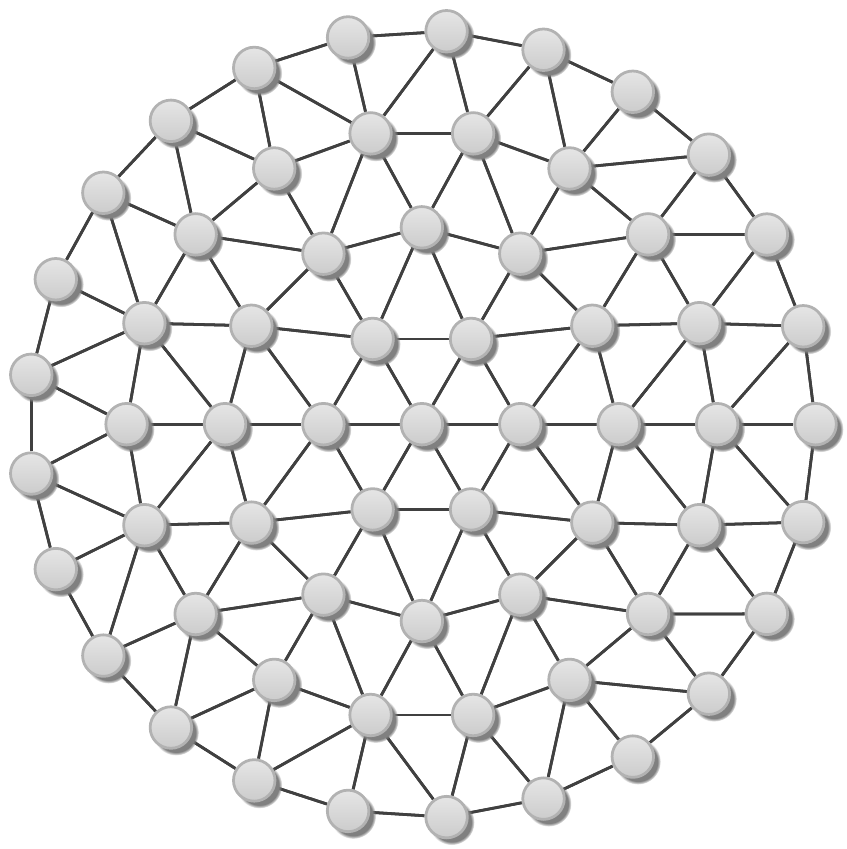}%
  \caption{Topology used for simulation.\label{fig:topology-62}}
\end{minipage}
\begin{minipage}{0.48\textwidth}
\centering
        \includegraphics[width=184.30pt]{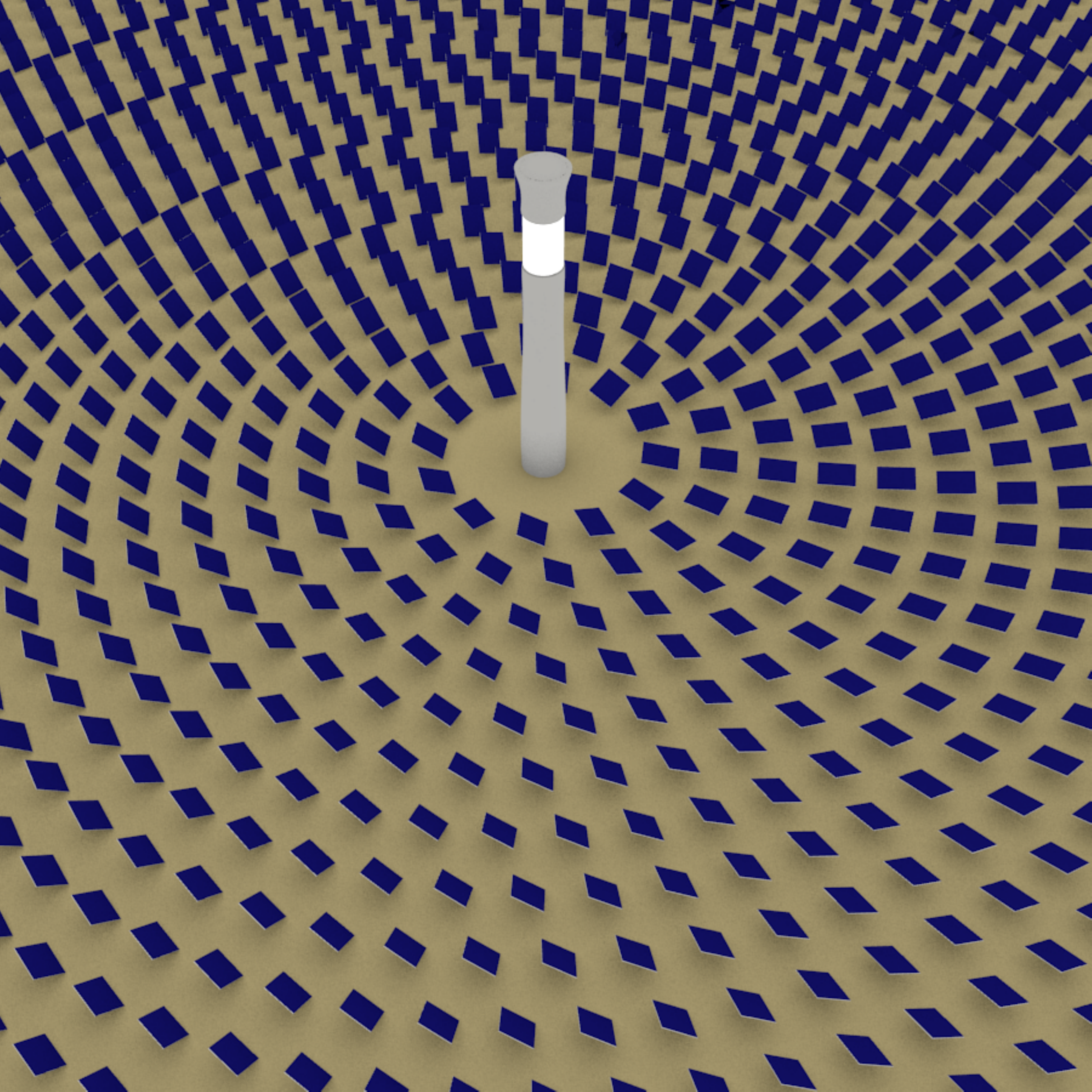}%
  \caption{Illustration of a solar tower power plant.\label{fig:solartower}}
\end{minipage}
\end{figure} %

\section{Simulative Evaluation}
\label{sect:simulation}
In order to demonstrate the performance of the implementation of openDSME and the presented approach for managing slots, this section presents the results of a simulative evaluation using the OMNeT++ simulation environment \cite{varga2001omnet++} together with the INET framework. 
As topology a regular network of concentric circles as shown in Fig.~\ref{fig:topology-62} is chosen. It resembles the previously mentioned application of a wireless network in a solar tower power plant \cite{pfahl_holistic_2014} as illustrated in Fig.~\ref{fig:solartower}. GPSR \cite{karp_gpsr:_2000} is used as routing protocol since it promises high scalability and low overhead. In contrast to the original greedy selection of the next hop, we use a variant of GPSR that minimizes the distance to the straight line between the original sender and the sink as proposed by \cite{lubkert_joint_2015} to significantly increase the scalability of GPSR.

Every node generates packets with the maximum payload of $127\,\text{Bytes}$ to be sent to the central node. The packets are generated either with a fixed interval of $\Iup$ or according to a Poisson distribution with mean $\Iup$. This data is then routed via GPSR to the central node where the received messages are counted and duplicates are discarded by using sequence numbers. Since we are interested in the results in the steady state and want to avoid effects originating from starting and stopping the simulation, the following procedure is applied:
\begin{enumerate}
\item Traffic generation starts with the simulation, but packets are marked as warm-up traffic.
\item After a setup time of $\tsetup=15\,\text{min}$, the generated packets are flagged as measurement packets.
\item When $\numpackets=100$ packets are sent, the generated packets are flagged as cool-down packets. If no measurement packet was received for $\tcooldown=15\,\text{s}$, the run is stopped.
\end{enumerate}

The end-to-end packet delivery ratio $\PDRi$ for node $i$ is calculated from the number of received packets $\Nr{i}$ at the central node as $\PDRi = \frac{\Nr{i}}{\numpackets}$. In the following figures, the PDR is shown as the mean over all $\PDRi$ together with the $95\%$ confidence intervals for this mean for 10 runs.

For this evaluation, DSME is compared to conventional CSMA/CA with acknowledgments and retransmissions as described in the IEEE 802.15.4 standard. In order to allow for a fair comparison, a pre-evaluation of CSMA/CA was conducted to find the optimal parameters for this scenario as presented in Appendix \ref{sect:csmapreevaluation}.

\subsection{Reliability}
\label{sect:reliability}

\begin{figure}[b]
\centering
\begin{subfigure}[t]{0.49\textwidth}
\centering
        \includegraphics[width=194.87pt]{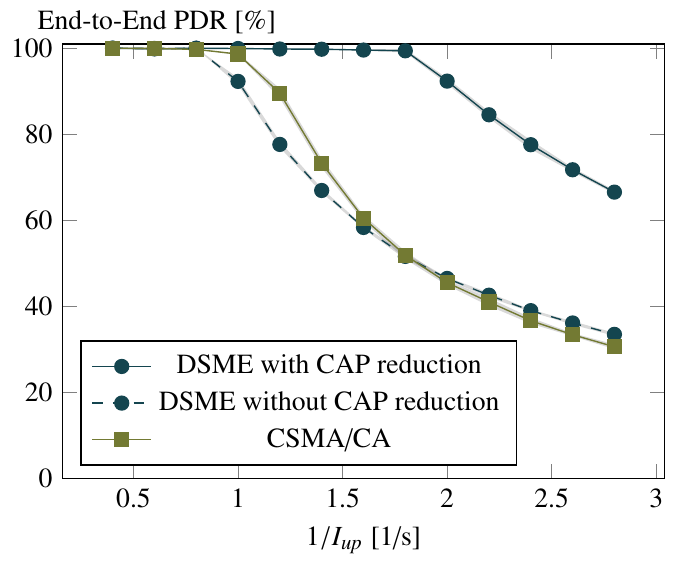}%
        \caption{Poisson Traffic\label{fig:csmadsme-pdr-Poisson}}
\end{subfigure}
\begin{subfigure}[t]{0.49\textwidth}
\centering
        \includegraphics[width=194.87pt]{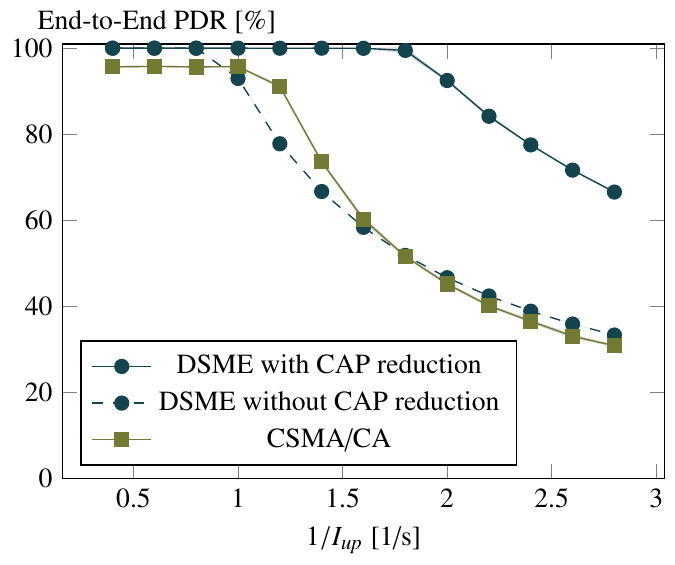}%
        \caption{Fixed Generation Interval\label{fig:csmadsme-pdr-FixedInterval}}
\end{subfigure}
\caption{Comparison of the end-to-end packet delivery ratio of CSMA/CA and DSME.\label{fig:csmadsme-pdr}}
\end{figure} %

In Fig.~\ref{fig:csmadsme-pdr}, the packet delivery ratios (PDR) of DSME with $\MO=6$ and CSMA/CA are compared. For the Poisson traffic generation in Fig.~\ref{fig:csmadsme-pdr-Poisson}, the traffic can successfully be delivered for a packet sending rate of $0.8\,\text{Hz}$ or lower for all scenarios. For higher rates the PDR decreases due to the increasing amount of traffic in the network. Without CAP reduction, less than half of the time is used to transmit data packets (the other time is reserved for beacons and the CAP). Still, the performance is nearly as good as CSMA/CA that suffers from collisions for rates over $0.8\,\text{Hz}$. DSME with CAP reduction, however, can still deliver the packets for twice as much traffic of $1.6\,\text{Hz}$, where CSMA/CA only delivers about $61\%$ of the packets. It is obvious that the higher complexity of DSME compared to CSMA/CA pays of by providing a much higher reliability.

\begin{figure}[tb]
\centering
        \includegraphics[width=353.27pt]{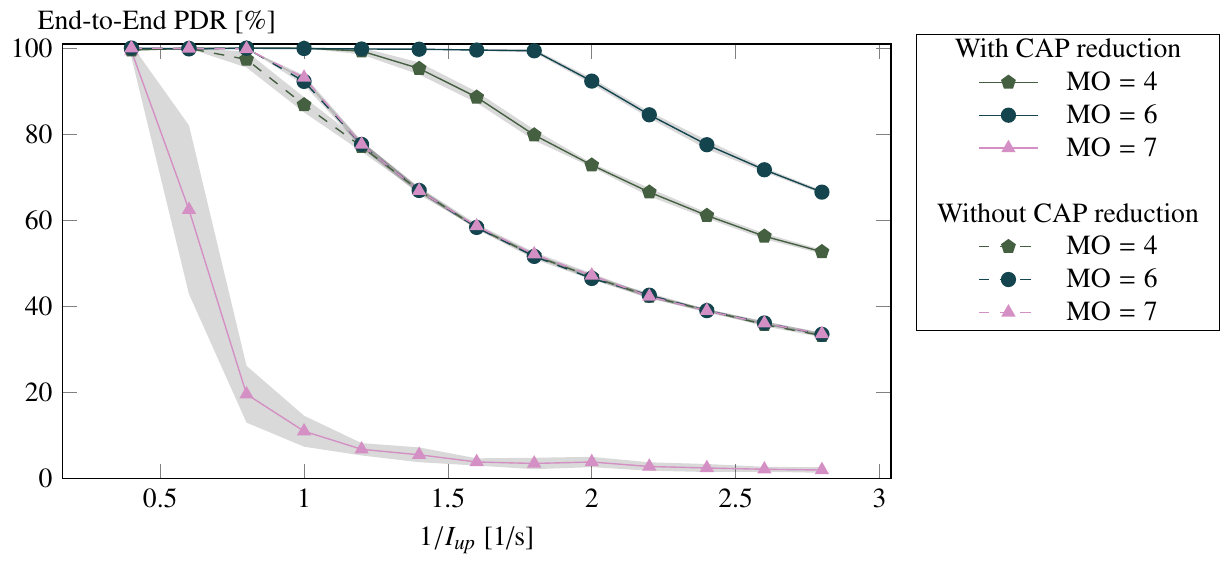}%
\caption{Packet delivery ratio for different DSME settings.\label{fig:dsme-pdr}}
\end{figure} %

When comparing these results with those of the fixed generation interval in Fig.~\ref{fig:csmadsme-pdr-FixedInterval}, we see that the PDR of CSMA/CA never goes above $96\%$. 
This can be explained by the fact that all nodes generate the packets at the same time so the probability of collisions with CSMA/CA is highly increased. While it is quite common for sensors in industrial applications to generate sensor readings at synchronized points in time, this should be avoided when using CSMA/CA or at least mitigated by additional artificial delays.
For DSME, the distribution has no influence on the results since packet collisions are avoided.

Fig.~\ref{fig:dsme-pdr} demonstrates the influence of the $\MO$ parameter. Only the results for the Poisson traffic generation are shown since the differences are not significant. Without CAP reduction, the influence of the $\MO$ parameter is small as it only controls the granularity of the schedule. Therefore, the corresponding graphs are grouped closely. With CAP reduction, however, the effect is significant since it determines the overall share of the CAP thus increases the number of GTS per time. By means of equations (\ref{eq:gtsnocapred}) and (\ref{eq:gtscapred}), the share of the CFP in a multi-superframe is calculated as approximately $69\%$ for $\MO=4$, $88\%$ for $\MO=6$ and $91\%$ for $\MO=7$. Therefore, the throughput, and with it the PDR, increases with increasing $\MO$. However, since the share of the CAP is reduced to only about $3\%$ for $\MO=7$ and at the same time, more GTS handshakes are required to allocate the increasing number of GTS, the CAP can no longer handle the amount of management traffic for $\MO=7$. So the PDR drops early and sharply since the nodes are not longer able to acquire enough slots in time. Lowering $\ap$ and increasing the warm-up period can mitigate this problem at the cost of a less reactive network.

\subsection{Delay}
\begin{figure}[t]
\centering
\begin{subfigure}[t]{0.48\textwidth}
\centering
        \includegraphics[width=185.09pt]{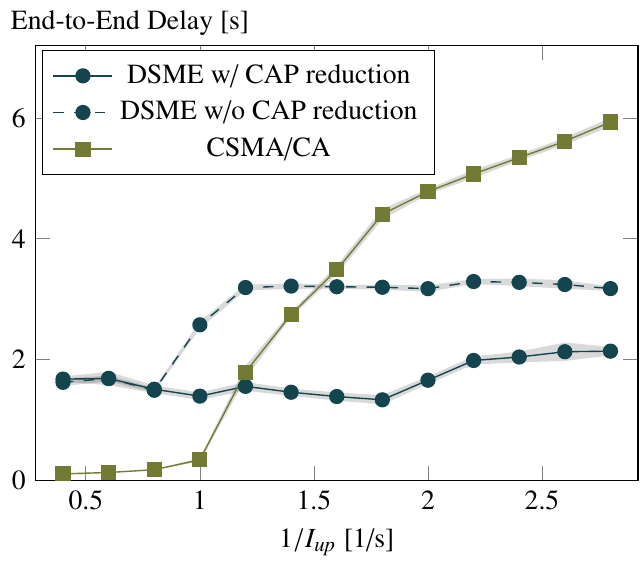}%
\caption{End-to-end delay over the generation rate.\label{fig:csmadsme-delay-rate-poisson}}
\end{subfigure}
\quad
\begin{subfigure}[t]{0.48\textwidth}
\centering
        \includegraphics[width=185.09pt]{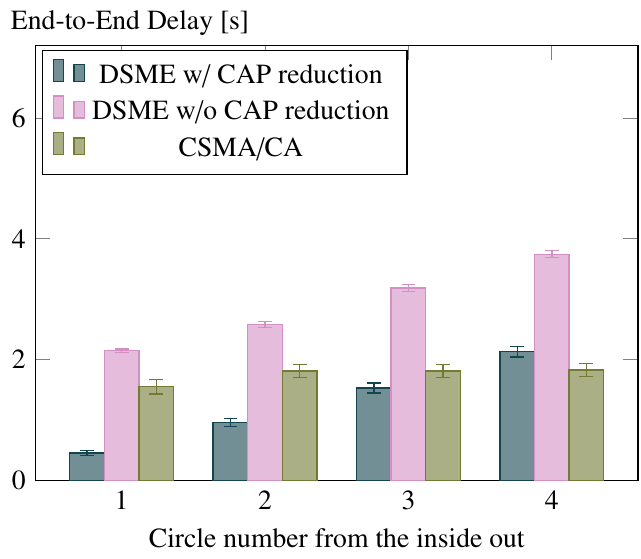}%
    \caption{End-to-end delay for every circle, corresponding to the number of hops, for $\Iup = \tfrac{1}{1.2} \,\text{s}$.\label{fig:circle-delay-poisson}}
\end{subfigure}
\caption{Comparison of the average end-to-end delay of CSMA/CA and DSME for Poisson traffic generation.\label{fig:csmadsme-delay-poisson}}
\end{figure}

\begin{figure}[b]
\centering
\begin{subfigure}[t]{0.48\textwidth}
\centering
        \includegraphics[width=185.09pt]{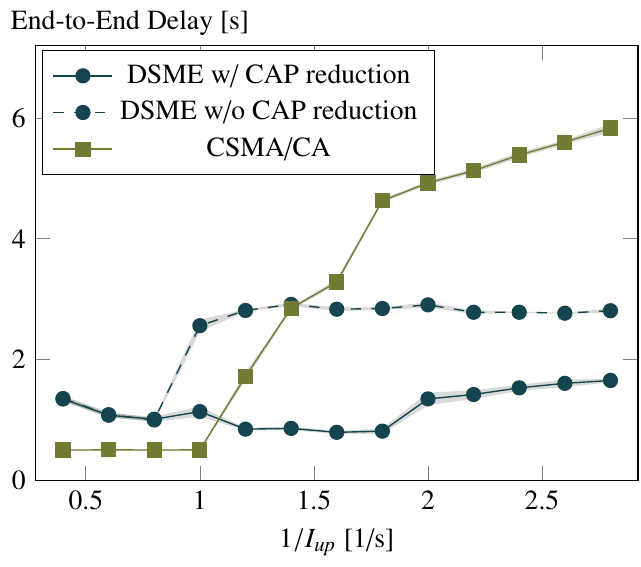}%
\caption{End-to-end delay over the generation rate.\label{fig:csmadsme-delay-rate-fixed}}
\end{subfigure}
\quad
\begin{subfigure}[t]{0.48\textwidth}
\centering
        \includegraphics[width=185.09pt]{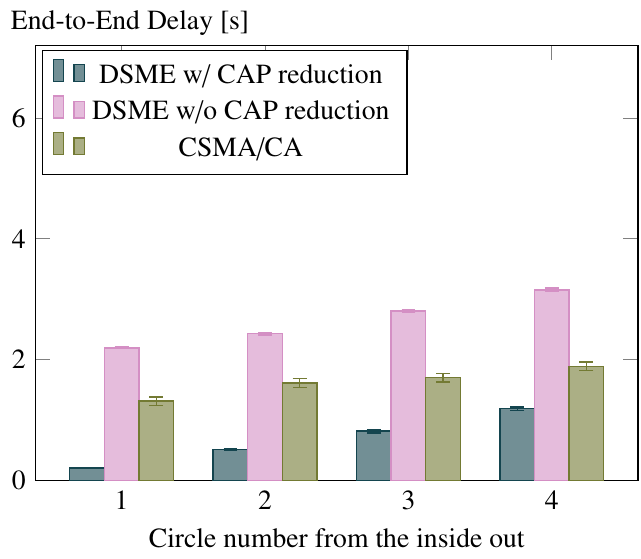}%
    \caption{End-to-end delay for every circle, corresponding to the number of hops, for $\Iup = \tfrac{1}{1.2} \,\text{s}$.\label{fig:circle-delay-fixed}}
\end{subfigure}
\caption{Comparison of the average end-to-end delay of CSMA/CA and DSME for fixed generation intervals.\label{fig:csmadsme-delay-fixed}}
\end{figure}

Many industrial applications do not only come with strict reliability constraints, but also require timely delivery of data packets. Fig.~\ref{fig:csmadsme-delay-rate-poisson} compares the average end-to-end delay of the packet delivery. In this plot, packets that are not delivered are not included in the statistic. Clearly, DSME has a disadvantage compared to CSMA/CA for this metric at low data rates since packets are delayed until the next matching transmission slot. This waiting time is often much longer than the backoff of CSMA/CA. In our experiments, this is due to our scheduling algorithm that does not try to optimize the delay, so it is expected that this leaves a lot of room for improvements. For larger data rates, however, where CSMA/CA is no longer able to deliver all packets and even those that get through face a much higher delay. This is due to retransmissions and queuing so that using DSME is advantageous again. Also in DSME the queuing delay increases when the network gets saturated. Without CAP reduction, this effect can be seen at around $1\,\text{Hz}$.
The corresponding plot for the fixed packet generation intervals, depicted in Fig.~\ref{fig:csmadsme-delay-rate-fixed}, shows a lower delay for DSME due to the lower variance of the traffic, but the qualitative statements still hold.

For looking at this in more detail, Fig.~\ref{fig:circle-delay-poisson} and Fig.~\ref{fig:circle-delay-fixed} distinguish between the different circles of the network for a rate of $1.2$ packets per second. Obviously, the end-to-end delay increases with the number of hops. Since the DSME network with CAP reduction is not saturated, the additional delay per hop is approximately constant. Without CAP reduction, the network is saturated in the center at this packet rate so the delay is significantly higher already for the innermost circle. This indicates that the packets fill up the queues of the nodes in the innermost circle, while the outer circles are not yet saturated. For the saturated CSMA/CA scenario, the largest part of the delay occurs on the innermost circle, too, while the differences of the outer circles are even within the confidence intervals, indicating that the network is not saturated in the outer cycles, again.

\subsection{GTS Management}

\begin{figure}[b]
    \vspace{-0.35cm}
\centering
\begin{subfigure}[t]{0.49\textwidth}
\centering
        \includegraphics[width=186.73pt]{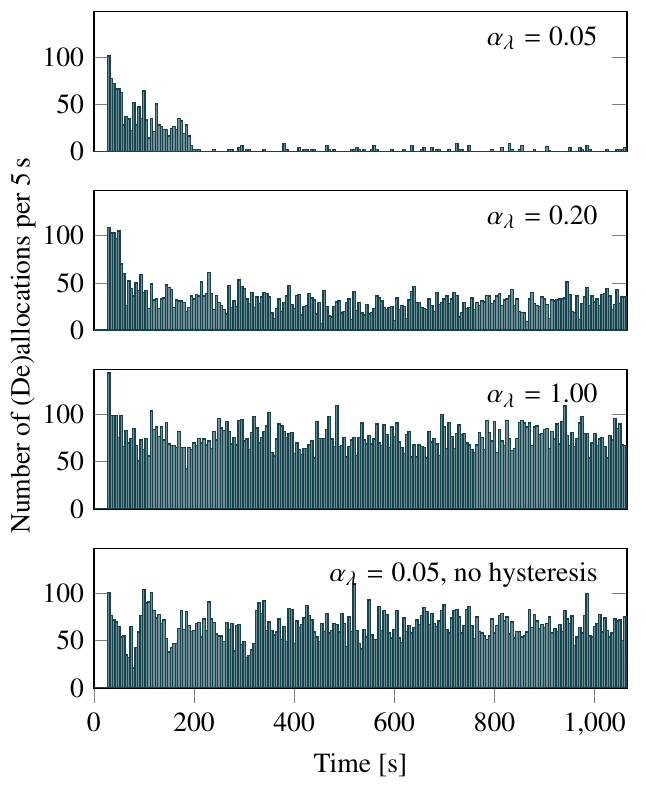}%
        \caption{Poisson Traffic\label{fig:tps-poisson}}
\end{subfigure}
\begin{subfigure}[t]{0.49\textwidth}
\centering
        \includegraphics[width=180.29pt]{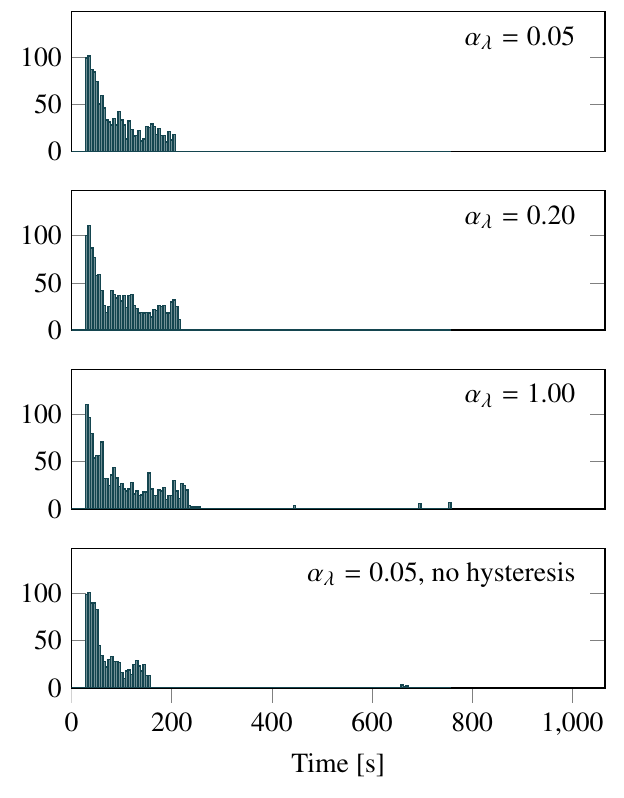}%
        \caption{Fixed Generation Interval\label{fig:tps-fixed}}
\end{subfigure}
  \caption{Number of allocations and deallocations per $5\,\text{s}$ over time for different settings of the proposed scheduling mechanism. \label{fig:tps}}
\end{figure} %

As already mentioned before, the influence of the management traffic can not be neglected. Though it takes place in the CAP and therefore does not directly influence the transmission in the CFP, a smooth GTS management is required for a efficient and error-free schedule. If the CAP is congested, GTS handshakes can not be performed in time as required (cf. $\MO=7$ with CAP reduction in Fig.~\ref{fig:dsme-pdr}). Furthermore, errors during the GTS handshake can lead to an inconsistent schedule, even though the procedure tries to mitigate these (for details see \cite{formaldsme}).

Fig.~\ref{fig:tps} depicts the influence of the proposed slot management on the amount of allocations and deallocations over time for $\Iup = 1\,\text{s}$. For every $5\,\text{s}$, the total number of GTS handshakes in the network is summed up. With the value of $\ap = 0.05$ we see an initialization phase until about $200\,\text{s}$ in Fig.~\ref{fig:tps-poisson}. In this phase, the nodes associate to the network and reserve the required amount of slots along the multi-hop paths. After that only very few changes in the schedule take place as predicted in Sect.~\ref{sect:alpha}.
With increasing $\ap$, the amount of management traffic increases to account for the higher volatility of the traffic prediction. This can lead to a better adaption of the current requirements, but at the previously mentioned costs. For $\ap = 1$, that is without any smoothing, the management traffic does not significantly decrease over time and no stable state can be reached. The same holds for the last scenario where $\ap=0.05$, but no hysteresis is applied (i.e. $\creq = \left\lceil\rateint\right\rceil$). Note that the first bars match for the scenario with and without hysteresis, because it is not relevant when $\rateint$ is much smaller than $\pt$.
For a fixed generation interval as shown in Fig.~\ref{fig:tps-fixed}, $\ap$ has a negligible influence since after the initialization phase, the traffic and thus the required number of slots over every link is constant and no further (de)allocations are required.

\section{Hardware Evaluation}
\label{sect:hardware}

In this section, a multi-hop hardware evaluation in the FIT IoT-LAB testbed \cite{iotlab} is presented. The main purpose of this evaluation is to demonstrate that openDSME is in fact suitable for execution on wireless nodes as well as to present some properties that are best shown on a real system, such as power consumption. The hardware, called M3OpenNode, consists of an AT86RF231 radio chip from Microchip/Atmel and an ARM Cortex M3 STM32F103REY by STMicroelectronics.

As operating system, Contiki with 6LoWPAN stack is used where the CSMA MAC is replaced by openDSME. This entails the use of RPL for routing. Since we target maximum reliability and stable network conditions, RPL is configured to use the Minimum Rank with Hysteresis Objective Function (MRHOF) with ETX metric.

As shown in \cite{kauer_constructing_2018}, constructing multi-hop topologies in dense wireless networks such as the FIT IoT-LAB can be difficult, because the physical distance between nodes correlates only slightly with the path loss between these nodes, due to effects such as reflection or antenna alignment. Therefore, the approach presented in \cite{kauer_constructing_2018} is used to generate line and tree topology used in this evaluation.

\subsection{Schedule}
\begin{figure}[b]
\centering
        \includegraphics[width=229.42pt]{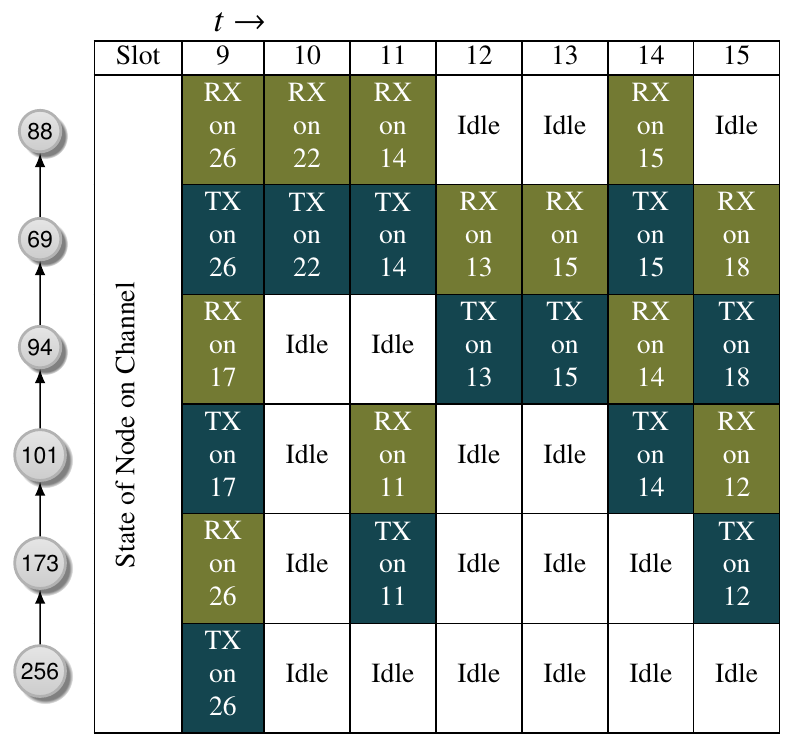}%
\caption{Line topology with associated emerging schedule in the hardware experiment.\label{fig:linetopology}}
\end{figure} %

In Fig.~\ref{fig:linetopology} a line topology with 6 nodes in the Lille testbed is shown together with an emerging schedule. Similar to the simulative evaluation, every node (apart from the sink 88) sends packets on average every $500\,\text{ms}$ towards the sink. As before, the nodes allocate slots based on the amount of traffic over the link. No CAP reduction is applied and $\SO=\MO=4$, so there is only one superframe per multi-superframe with a duration of
$2^{4}\cdot 960\,\text{symbols}\cdot 16\,\frac{\mu\text{s}}{\text{symbol}} = 245.8\,\text{ms}$
according to (\ref{math:SD}). So one slot is sufficient for transmitting on average about 4 packets per second. When applying the hysteresis to account for times of momentarily higher traffic, the shown schedule emerges. The higher the number of nodes that route over a given link, the more slots are allocated. It is also apparent that slots are reused. For example, time slot 9 (the first slot in the CFP), is used on channel 26 for the link $69 \to 88$, while it is used at the same time on channel 17 for link $101 \to 94$ as well as for $256 \to 173$, again on 26. This is valid, since the transmission from 256 is out of range for node 88.

\subsection{Network Formation}
\label{sect:formation}

As explained in Sect.~\ref{sect:timesync}, the nodes first need to synchronize to a beacon, then associate to the network. In the second phase, they potentially become coordinators on their own to pass on the time synchronization. In openDSME, the following procedure is implemented to maintain an even distribution of the coordinators. An associated node that is not a coordinator (so far), will count the number of coordinators it recognizes in the neighborhood. If less than two beacons were received, it randomly decides to become coordinator itself with a chance of $\sfrac{1}{3}$ every beacon interval. Also, when a node was performing a passive scan for too long without receiving any beacon, it will switch to active scan and send out a beacon request. A node receiving a beacon request will also turn into a coordinator. This procedure minimizes the number of coordinators and allows for very dense networks while lowering the energy consumption and ensuring a connected network.

\begin{figure}[hb]
\centering
        \includegraphics[width=240.84pt]{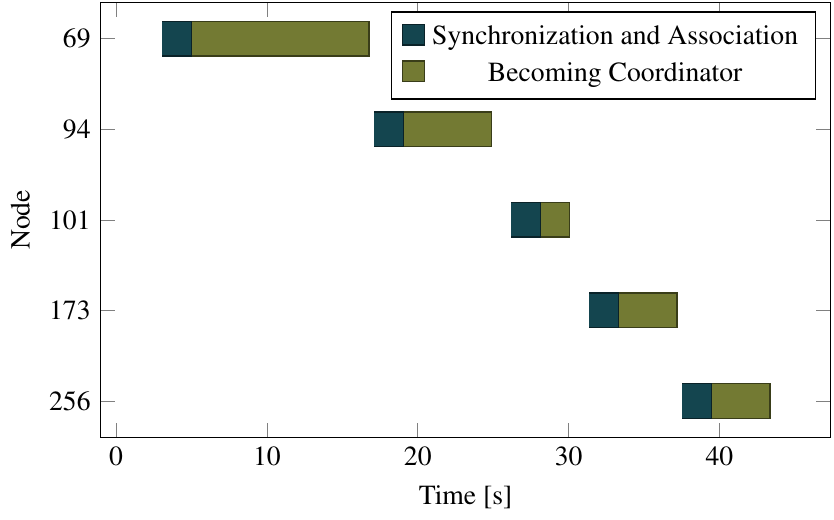}%
\caption{Network formation in the FIT IoT-LAB line topology experiment.\label{fig:timeline}}
\end{figure}

In the shown line topology, all nodes have to become coordinators to enable the functioning of the network. Fig.~\ref{fig:timeline} demonstrates the associated timing.
Since the parent of a node has to become coordinator before its child can synchronize to it, the shown waterfall shape emerges. After about 40 seconds all nodes are associated. This time is influenced by several parameters such as the scanning duration and the beacon interval. 

\subsection{Energy Consumption}
Besides the high reliability as demonstrated in Sect.~\ref{sect:simulation}, DSME has the advantage of consuming less energy than an always-on CSMA/CA MAC, because the transceiver can be turned off during unused slots in the CFP. While it is not explicitly targeted to consume as few power as possible, such as low-power listening approaches \cite{polastre_versatile_2004}, the consumption can be reduced significantly as shown in this section.

\begin{figure}[htb]
\begin{subfigure}[t]{0.49\textwidth}
\centering
        \includegraphics[width=127.04pt]{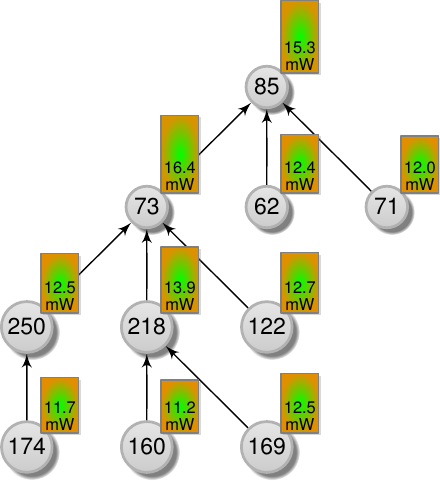}%
\caption{Tree topology and results for a single scenario.\label{fig:treetopology}}
\end{subfigure}%
\begin{subfigure}[t]{0.51\textwidth}
\centering
      \includegraphics[width=188.82pt]{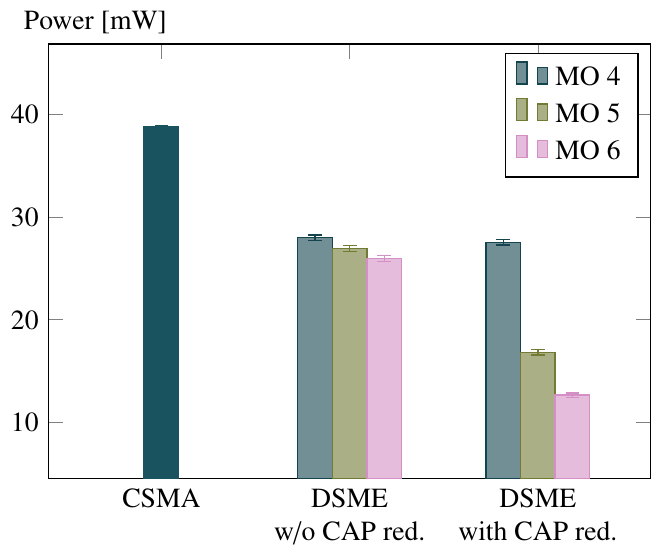}%
  \caption{Measured power consumption for different scenarios.\label{fig:power}}
\end{subfigure}%
  \caption{Power consumption experiment in the FIT IoT-LAB.}
\end{figure} %

For this experiment, a tree topology with 10 nodes is built. Every node sends a packet on average every 2 seconds to the PAN coordinator. Fig.~\ref{fig:treetopology} shows one resulting RPL routing tree, annotated with the average power consumption of each node for an exemplary run with $\SO=4$, $\MO=6$ and enabled CAP reduction. Of course, since RPL is a dynamic routing and the environmental conditions can change, the routing tree is different for other runs. The stated power is the additional power used by the transceiver, excluding the power for the CPU and other peripherals. For getting these values, the idle consumption of each the M3OpenNodes as published in \cite{iotlab_consumption} is subtracted from the total power consumption. While the figure only represents a single run and the results are thus quite diverse, it is already apparent that the nodes with a high number of connections have a higher power consumption due to the higher number of reserved slots.

In order to get more significant results, this experiment was repeated for various MO settings, with and without CAP reduction as well as for an always-on CSMA/CA MAC and every setup was repeated ten times. Fig.~\ref{fig:power} shows the resulting power consumptions, averaged over the network, together with the $95\%$ confidence intervals. The power consumption in the DSME experiments without CAP reduction is about $10\,\text{mW}$ smaller than the one for the CSMA experiment. However, the influence of MO is small and can be explained by the higher granularity. For $MO=4$, the nodes consume the same amount of power, regardless of the CAP reduction, because there is only one superframe per multi-superframe anyway. Though, the skipped CAP for the other MO settings leads to a significant reduction in power consumption.

\section{Open Issues}
\label{sect:issues}

The management procedure described in the DSME standard aims to avoid all slot collisions per design and handles more corner cases than all scheduling approaches for TSCH known to the authors, especially with regard to hidden node situations. Still it has some issues in special cases as outlined in the following that need to be solved for a fully robust slot management.

\subsection{Removing Occupied Slots from SAB}
\begin{figure}[t]
\centering
        \includegraphics[width=220.49pt]{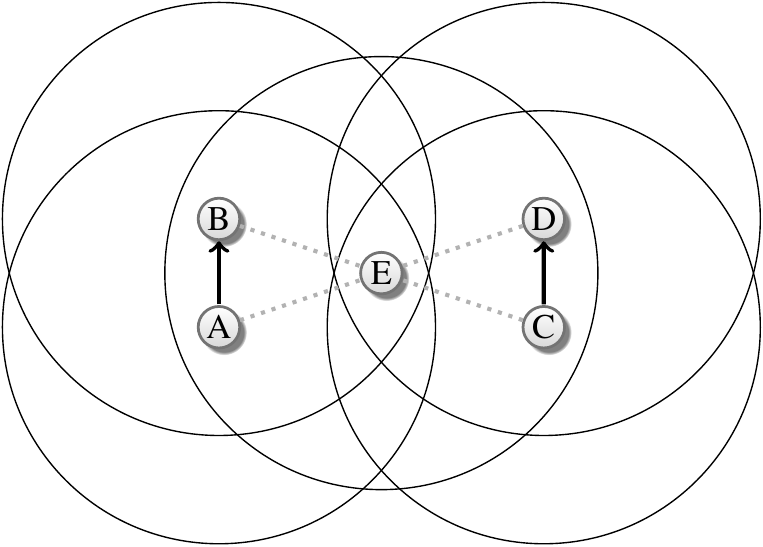}%
\caption{Situation where a slot might be incorrectly removed from the SAB.\label{fig:remsab}}
\end{figure}

One situation that can lead to unwanted results is depicted in Fig.~\ref{fig:remsab}. In this situation two pairs of devices A/B and C/D use the same time and frequency slot. Since they are not in range of each other, this is no problem as such and even improves spatial reuse. Node E was informed during the slot allocation handshake that this slot shall not be used.

If now for example C decides to deallocate this slot, they perform a deallocation handshake. By doing so they inform node E about the deallocation. Since node E does not store the information how many neighbors reserved the slot, it will be marked as free, leading to an inconsistent slot allocation bitmap (SAB).

A solution would be to use a counter instead of a single bit, but since the situation is assumed to occur rarely and B will send a duplicate allocation message anyway if E tries to use the slot, this is not worth the additionally required memory.

\subsection{Eternal Blocking of Slots}
A more severe problem that is especially relevant for long running networks with a volatile topology is described in the following. There are multiple variants of this problem, but all have in common that slots are not removed from the SAB even though no pair of nodes is still using it. Over time, this will fill up the SAB until no more free slots are available and no new allocation can take place.

This happens for example if the radio conditions between two nodes deteriorate significantly. Even if this is recognized by the nodes (see Sect. \ref{sect:depre}), they are no longer able to perform a proper deallocation handshake. In particular, no response and no notify message will be sent to inform the neighbors that the slot is no longer in use.

The same situation will occur in case of mobility or if a battery powered node runs out of energy and shuts down before allocated slots can be deallocated. While the latter could be mitigated by storing the ACT in persistent memory, a general solution is in demand to avoid the eternal blocking of slots.

A possible solution might be to remove occupied slots from the SAB that were not modified for a long time, indicated by a single additional bit per slot. In regular intervals of maybe about an hour, this flag will be set for every occupied slot. If no deallocation took place after another hour, the bit is still set and the slot will be marked as free in the SAB. Of course, as for the previous issue, one has to rely on the duplicate allocation message to avoid inconsistencies in this case. This could be improved by leasing slots only for a certain time span after which they have to be reallocated.

\subsection{Late Removal of Invalid Slots}
The slot management handshake is performed during the contention access phase (CAP). Therefore, collisions and thus message loss are quite common in dense networks with many allocations and deallocations. In case the notify is lost, only one of the nodes of a link will write the slot to the allocation counter table (ACT). This problem is analyzed in depth in \cite{formaldsme}, including a proposal to solve this problem by adding an additional flag to the ACT.

\section{Conclusion}
\label{sect:conclusion}

Reliable wireless networks are an integral component in the realization of the Industrial Internet of Things (IIoT). The Deterministic and Synchronous Multi-channel Extension (DSME) provides collision-free time-slotted communication in IEEE 802.15.4 networks and is thus a promising candidate for flexible and energy-efficient connection of sensors and actuators to the IIoT.
This paper presents a comprehensive analysis of the DSME as well as the open-source implementation openDSME. Furthermore, a technique for traffic-aware and decentralized slot management is proposed and evaluated. DSME aligns the transmissions in time and frequency slots and by this aims for a much higher reliability than the conventional CSMA/CA medium access. By using a slot allocation handshake that is overheard by all neighbors, slot collisions can be avoided. 
The results show that in fact DSME can reliably deliver data packets in a large multi-hop topology for twice the amount of traffic than CSMA/CA.
An extensive analysis of the relevant parameters is presented and the evaluation demonstrates the pros and cons of the CAP reduction.

Hardware experiments in the FIT/IoT-LAB demonstrates the applicability of openDSME for a physically deployed network and show the reduced energy consumption of IEEE 802.15.4 due to the possibility to disable the transceiver during unused slots. The presented DSME implementation works seamlessly together with the 6LoWPAN stack of Contiki including RPL without any further support by the stack.

Finally, open issues are presented that encourage future work on the topic of distributed slot management. Developing scheduling techniques that reduce the end-to-end delay, while maintaining the traffic-awareness and scalability of the proposed technique would make the use of DSME in the Industrial Internet of Things even more promising.

\begin{acks}
The authors would like to thank everyone who has contributed to the development of openDSME, especially Tobias L\"ubkert for the first functional OMNeT++ DSME implementation, Sandrina K\"ostler for mastering the complex data structures, Axel Neuser for the Contiki port and Florian Meyer for the channel hopping and CAP reduction functionality.
This research was partially supported by the German Federal Ministry for Economic Affairs and Energy (BMWi) via the AutoR Project (0325629D).
\end{acks}

\appendix

\section{Calculations for Analyzing the EWMA Filter}
\label{sect:ewmacalculations}

Algorithm~\ref{algo:alpha} is applied to calculate an upper bound $\Pintupper$ and a lower bound $\Pintlower$ for $\Pint$. This algorithm requires a probability mass function with $\tilde{f}_X(m) = 0\,\,\forall m > \mmax$ for a fixed $\mmax \gg \mu$. This is a minor constraint for most practical applications while popular generic distributions, such as the Poisson distribution, can be approximated by such a probability distribution by applying
\begin{align}
\tilde{f}_X(m) = \begin{cases}
f_X(m) & m < \mmax\\
\sum_{k = \mmax}^{\infty} f_X(k) & m = \mmax\\
    0 & \text{else.}
\end{cases}
\end{align}

Secondly, to bypass the fact that the number of elements in $K_n$ grows exponentially, the interval from $0$ to $\mmax$ is split into $N=\mmax\cdot h$ intervals of size $\sfrac{1}{h}$ with a chosen integer $h > 1$. A larger $h$ will lead to tighter bounds and a longer execution time. Also a small $\varepsilon > 0$ is to be chosen. 

\begin{algorithm}[tbh]
  \DontPrintSemicolon
  \SetKwFunction{FStep}{Step}
  \SetKwFunction{FIntervalProbability}{IntervalProbability}
  \SetKwProg{Fn}{function}{}{\Return{$y_n$}}
  \Fn{\FStep{$y_{n-1},v$}}{
    $y_n \leftarrow \left[0,0,0,\ldots,0\right]$, with $\left|y_n\right| = N$\;
    \For{$i \leftarrow 0,\ldots,N-1$}{
      \For{$m \leftarrow 0,\ldots,\mmax$}{
            $\lambda \leftarrow m\cdot\ap + \left(\sfrac{i}{h}+v\right)\cdot(1-\ap)$\;
            $p \leftarrow \tilde{f}_X(m) \cdot y_{n-1}(i)$ \;
            $s \leftarrow \min\!\left(\left\lfloor\lambda \cdot h\right\rfloor, N-1\right)$ \;
            $y_n(s) \leftarrow y_n(s) + p$\;
      }
    }
  }
  \;
  \SetKwRepeat{Do}{do}{while}
  \SetKwProg{Fn}{function}{}{\Return{$\Pintupper,\Pintlower$}}
  \Fn{\FIntervalProbability}{
      $\check{y}_0 \leftarrow \left[1,0,0,\ldots,0\right]$, with $\left|\check{y}_0\right| = N$\;
      $\hat{y}_0 \leftarrow \left[0,0,0,\ldots,1\right]$, with $\left|\hat{y}_0\right| = N$\;
      $n \leftarrow 0$\;
      \Do{$\Pintupper-\Pintlower > \varepsilon$}{
        $n \leftarrow n + 1$\;
        $\check{y}_n \leftarrow$ \Call{Step}{$\check{y}_{n-1},0$}\;
        $\hat{y}_n \leftarrow$ \Call{Step}{$\hat{y}_{n-1},\sfrac{1}{h}$}\;
        $\Pintlower \leftarrow 1 - {\sum_{i = 0}^{\left\lceil\left(\mu - 1\right)\cdot h\right\rceil-1} \check{y}_n} - {\sum_{i = \left\lfloor\left(\mu + 1\right)\cdot h\right\rfloor}^{N}\hat{y}_n}$\;
        $\Pintupper \leftarrow 1 - {\sum_{i = 0}^{\left\lceil\left(\mu - 1\right)\cdot h\right\rceil-1} \hat{y}_n} - {\sum_{i = \left\lfloor\left(\mu + 1\right)\cdot h\right\rfloor}^{N}\check{y}_n}$\;
      }
  }
  \caption{Probability Bounds for $\mu-1 \leq \rateint \leq \mu+1$\label{algo:alpha}}
\end{algorithm}

For calculating the number of multi-superframes until $\mu-1$ is reached, the following calculation is applied. Given is the recursive equation
\begin{align*}
  \ratein{t} &= \mu \cdot \ap + \ratein{t-1} \cdot \left(1-\ap\right) \text{ with } \ratein{0} = 0.
\end{align*}
Solving for $\rateint$ as
\begin{align*}
\ratein{t} &= \left(1 - (1-\ap)^t\right) \cdot \mu.\\
\end{align*}
Setting $\ratein{t}$ equal to target $\mu-1$ and solving for $t$ finally leads to
\begin{align*}
\mu - 1 &= \left(1 - (1-\ap)^t\right) \cdot \mu\\
  t &= \frac{\log\left(\mu^{-1}\right)}{\log\left(1-\ap\right)}.
\end{align*}

\pagebreak
\section{CSMA/CA Pre-Evaluation}
\label{sect:csmapreevaluation}

\begin{figure}[tbhp]
\centering
    \includegraphics[width=334.56pt]{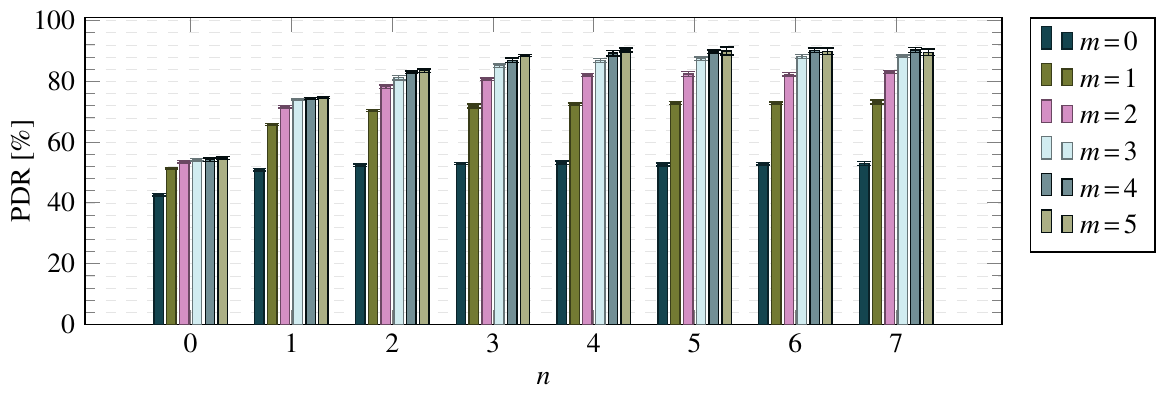}\\
    \includegraphics[width=236.30pt]{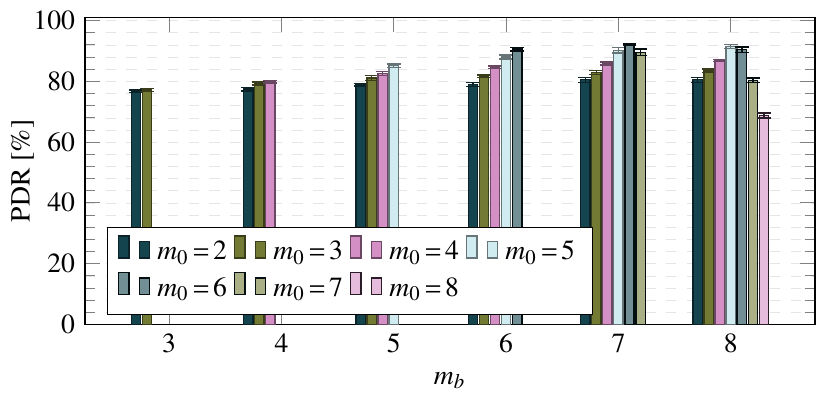}%
    \includegraphics[width=98.26pt]{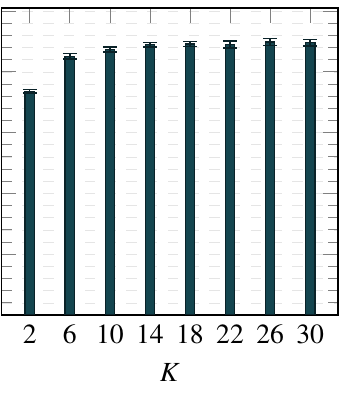}%
    \caption{Comparison of parameters for pure CSMA/CA. If not specified otherwise $\maxbackoffs=5$, $\maxretrans=7$, $\initbexp=\maxbexp=7$, $\maxqueue=30$.\label{fig:csma} and $\Iup = \tfrac{1}{1.2} \,\text{s}$ for Poisson traffic generation.\label{fig:csma-poisson}}
\end{figure}

\begin{figure}[tbhp]
\centering
    \includegraphics[width=334.56pt]{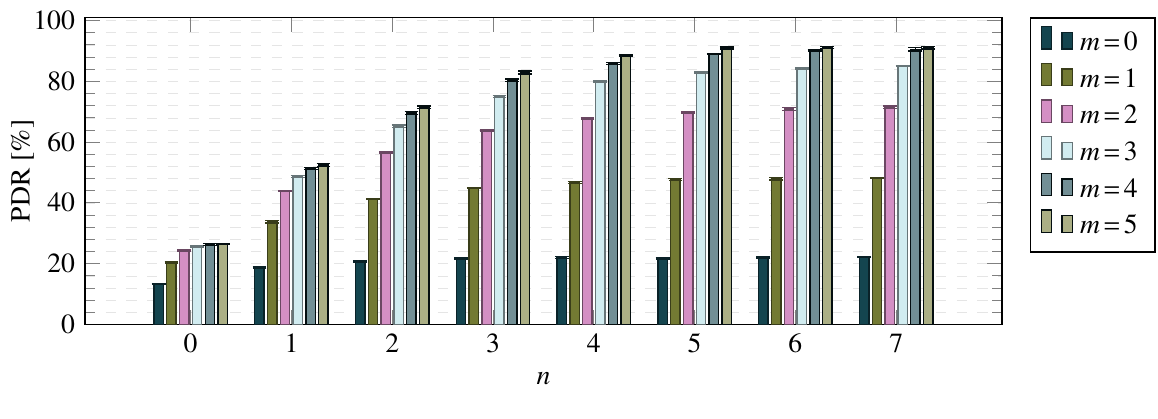}\\
	\includegraphics[width=236.30pt]{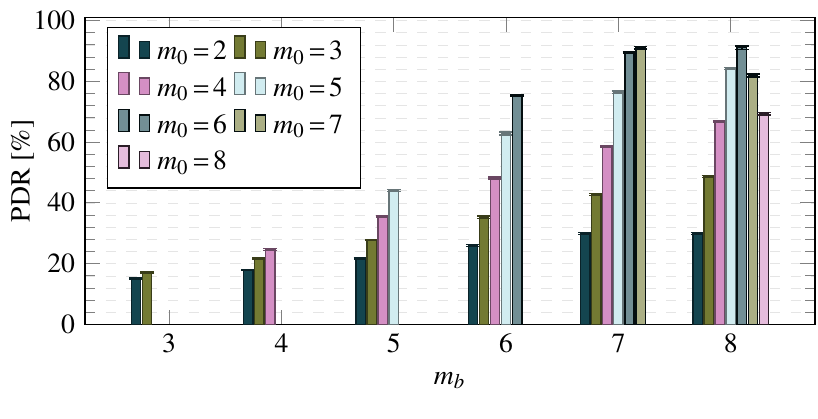}%
	\includegraphics[width=98.26pt]{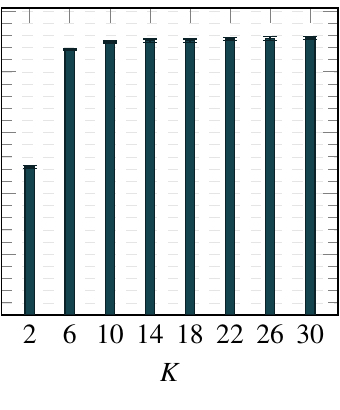}%
    \caption{Comparison of parameters for pure CSMA/CA. If not specified otherwise $\maxbackoffs=5$, $\maxretrans=7$, $\initbexp=\maxbexp=7$, $\maxqueue=30$.\label{fig:csma} and $\Iup = \tfrac{1}{1.2} \,\text{s}$ for fixed generation intervals.\label{fig:csma-fixed}}
\end{figure}

In this pre-evaluation, the most competitive parameters of CSMA/CA for the given scenario are determined. For Poisson traffic generation, the results are shown in Fig.~\ref{fig:csma-poisson}. The uppermost plot shows the results when changing \stdname{macMaxCsmaBackoffs} ($\maxbackoffs$) and \stdname{macMaxFrameRetries} ($\maxretrans$). In the given scenario, using the maximum parameters $\maxbackoffs=5$ and $\maxretrans=7$ results in the highest PDR. In general, retransmissions can improve the reliability, but also lead to a higher channel utilization and are less effective as generally assumed due to the simultaneous retransmission effect \cite{meiermodel}, therefore the gain for the higher number of retransmissions is negligible. The corresponding plot for the fixed generation intervals, Fig.~\ref{fig:csma-fixed}, shows the same trend, but the results are much worse for the lower parameters of $\maxbackoffs$ and $\maxretrans$ due to the synchronized packet generation as explained in Sect.~\ref{sect:reliability}.

When evaluating the values for \stdname{macMinBE} ($\initbexp$) and \stdname{macMaxBE} ($\maxbexp$), we see a tendency of higher PDRs towards higher parameter values, but the reliability decreases again for very large values. In the following we choose $\initbexp=\maxbexp=7$. Third, the queue length $\maxqueue$ is evaluated. Since DSME has a higher RAM overhead, the CSMA/CA stack could use a larger queue and still requires less RAM, while avoiding queue losses. In general, a longer queue increases the PDR, but only up to about $K=14$ for the given scenario. For larger values of $K$, the maximum queue length is rather irrelevant, because packets are mainly dropped due to collisions and not due to queue overflow.

This parameter selection may not be optimal for other scenarios, especially since they increase the congestion, delay and are not useful for volatile networks, but they are most competitive for the considered topology and traffic.

\bibliographystyle{ACM-Reference-Format}
\bibliography{reliable_multihop_dsme}


\begin{thebibliography}{50}


\ifx \showCODEN    \undefined \def \showCODEN     #1{\unskip}     \fi
\ifx \showDOI      \undefined \def \showDOI       #1{#1}\fi
\ifx \showISBNx    \undefined \def \showISBNx     #1{\unskip}     \fi
\ifx \showISBNxiii \undefined \def \showISBNxiii  #1{\unskip}     \fi
\ifx \showISSN     \undefined \def \showISSN      #1{\unskip}     \fi
\ifx \showLCCN     \undefined \def \showLCCN      #1{\unskip}     \fi
\ifx \shownote     \undefined \def \shownote      #1{#1}          \fi
\ifx \showarticletitle \undefined \def \showarticletitle #1{#1}   \fi
\ifx \showURL      \undefined \def \showURL       {\relax}        \fi
\providecommand\bibfield[2]{#2}
\providecommand\bibinfo[2]{#2}
\providecommand\natexlab[1]{#1}
\providecommand\showeprint[2][]{arXiv:#2}

\bibitem[\protect\citeauthoryear{Accettura, Vogli, Palattella, Grieco, Boggia,
  and Dohler}{Accettura et~al\mbox{.}}{2015}]%
        {accettura_decentralized_2015}
\bibfield{author}{\bibinfo{person}{N. Accettura}, \bibinfo{person}{E. Vogli},
  \bibinfo{person}{M.~R. Palattella}, \bibinfo{person}{L.~A. Grieco},
  \bibinfo{person}{G. Boggia}, {and} \bibinfo{person}{M. Dohler}.}
  \bibinfo{year}{2015}\natexlab{}.
\newblock \showarticletitle{Decentralized {Traffic} {Aware} {Scheduling} in
  6TiSCH {Networks}: {Design} and {Experimental} {Evaluation}}.
\newblock \bibinfo{journal}{\emph{IEEE Internet of Things Journal}}
  \bibinfo{volume}{2}, \bibinfo{number}{6} (\bibinfo{date}{Dec.}
  \bibinfo{year}{2015}), \bibinfo{pages}{455--470}.
\newblock
\showISSN{2327-4662}
\urldef\tempurl%
\url{https://doi.org/10.1109/JIOT.2015.2476915}
\showDOI{\tempurl}


\bibitem[\protect\citeauthoryear{Adjih, Baccelli, Fleury, Harter, Mitton, Noel,
  Pissard-Gibollet, Saint-Marcel, Schreiner, Vandaele, and Watteyne}{Adjih
  et~al\mbox{.}}{2015}]%
        {iotlab}
\bibfield{author}{\bibinfo{person}{C. Adjih}, \bibinfo{person}{E. Baccelli},
  \bibinfo{person}{E. Fleury}, \bibinfo{person}{G. Harter}, \bibinfo{person}{N.
  Mitton}, \bibinfo{person}{T. Noel}, \bibinfo{person}{R. Pissard-Gibollet},
  \bibinfo{person}{F. Saint-Marcel}, \bibinfo{person}{G. Schreiner},
  \bibinfo{person}{J. Vandaele}, {and} \bibinfo{person}{T. Watteyne}.}
  \bibinfo{year}{2015}\natexlab{}.
\newblock \showarticletitle{{FIT IoT-LAB: A Large Scale Open Experimental IoT
  Testbed}}. In \bibinfo{booktitle}{\emph{IEEE 2nd World Forum on Internet of
  Things (WF-IoT)}}.
\newblock
\urldef\tempurl%
\url{https://doi.org/10.1109/WF-IoT.2015.7389098}
\showDOI{\tempurl}


\bibitem[\protect\citeauthoryear{Alderisi, Patti, Mirabella, and
  Bello}{Alderisi et~al\mbox{.}}{2015}]%
        {alderisi_simulative_2015}
\bibfield{author}{\bibinfo{person}{Giuliana Alderisi}, \bibinfo{person}{Gaetano
  Patti}, \bibinfo{person}{Orazio Mirabella}, {and} \bibinfo{person}{Lucia~Lo
  Bello}.} \bibinfo{year}{2015}\natexlab{}.
\newblock \showarticletitle{Simulative Assessments of the {IEEE} 802.15. 4e
  {DSME} and {TSCH} in Realistic Process Automation Scenarios}. In
  \bibinfo{booktitle}{\emph{Proceedings of the 13th International Conference on
  Industrial {Informatics} ({INDIN} 2015)}}. \bibinfo{publisher}{IEEE},
  \bibinfo{pages}{948--955}.
\newblock
\urldef\tempurl%
\url{https://doi.org/10.1109/INDIN.2015.7281863}
\showDOI{\tempurl}


\bibitem[\protect\citeauthoryear{Alexander, Brandt, Vasseur, Hui, Pister,
  Thubert, Levis, Struik, Kelsey, and Winter}{Alexander et~al\mbox{.}}{2012}]%
        {rfc6550}
\bibfield{author}{\bibinfo{person}{Roger Alexander}, \bibinfo{person}{Anders
  Brandt}, \bibinfo{person}{JP Vasseur}, \bibinfo{person}{Jonathan Hui},
  \bibinfo{person}{Kris Pister}, \bibinfo{person}{Pascal Thubert},
  \bibinfo{person}{P Levis}, \bibinfo{person}{Rene Struik},
  \bibinfo{person}{Richard Kelsey}, {and} \bibinfo{person}{Tim Winter}.}
  \bibinfo{year}{2012}\natexlab{}.
\newblock \bibinfo{title}{{RPL: IPv6 Routing Protocol for Low-Power and Lossy
  Networks}}.
\newblock \bibinfo{howpublished}{RFC 6550}.
\newblock
\urldef\tempurl%
\url{https://doi.org/10.17487/RFC6550}
\showDOI{\tempurl}


\bibitem[\protect\citeauthoryear{Anamalamudi, Liu, Zhang, Sangi, Perkins, and
  Anand}{Anamalamudi et~al\mbox{.}}{2018}]%
        {sf1}
\bibfield{author}{\bibinfo{person}{Satish Anamalamudi}, \bibinfo{person}{Bing
  Liu}, \bibinfo{person}{Mingui Zhang}, \bibinfo{person}{Abdur Sangi},
  \bibinfo{person}{Charles Perkins}, {and} \bibinfo{person}{S.~V.~R. Anand}.}
  \bibinfo{year}{2018}\natexlab{}.
\newblock \bibinfo{title}{Scheduling {Function} {One} ({SF}1): Hop-by-Hop
  {Scheduling} with {RSVP}-{TE} in 6TiSCH {Networks}}.
\newblock
\newblock
\urldef\tempurl%
\url{https://tools.ietf.org/html/draft-satish-6tisch-6top-sf1-04}
\showURL{%
\tempurl}


\bibitem[\protect\citeauthoryear{Capone, Brama, Ricciato, Boggia, and
  Malvasi}{Capone et~al\mbox{.}}{2014}]%
        {capone_modeling_2014}
\bibfield{author}{\bibinfo{person}{S. Capone}, \bibinfo{person}{R. Brama},
  \bibinfo{person}{F. Ricciato}, \bibinfo{person}{G. Boggia}, {and}
  \bibinfo{person}{A. Malvasi}.} \bibinfo{year}{2014}\natexlab{}.
\newblock \showarticletitle{Modeling and Simulation of Energy Efficient
  Enhancements for {IEEE} 802.15.4e {DSME}}. In
  \bibinfo{booktitle}{\emph{Proceedings of the 2014 {Wireless}
  {Telecommunications} {Symposium}}}. \bibinfo{pages}{1--6}.
\newblock
\urldef\tempurl%
\url{https://doi.org/10.1109/WTS.2014.6835017}
\showDOI{\tempurl}


\bibitem[\protect\citeauthoryear{Civerchia, Bocchino, Salvadori, Rossi,
  Maggiani, and Petracca}{Civerchia et~al\mbox{.}}{2017}]%
        {civerchia_industrial_2017}
\bibfield{author}{\bibinfo{person}{Federico Civerchia},
  \bibinfo{person}{Stefano Bocchino}, \bibinfo{person}{Claudio Salvadori},
  \bibinfo{person}{Enrico Rossi}, \bibinfo{person}{Luca Maggiani}, {and}
  \bibinfo{person}{Matteo Petracca}.} \bibinfo{year}{2017}\natexlab{}.
\newblock \showarticletitle{Industrial {Internet} of {Things} Monitoring
  Solution for Advanced Predictive Maintenance Applications}.
\newblock \bibinfo{journal}{\emph{Journal of Industrial Information
  Integration}}  \bibinfo{volume}{7} (\bibinfo{date}{Sept.}
  \bibinfo{year}{2017}), \bibinfo{pages}{4--12}.
\newblock
\showISSN{2452414X}
\urldef\tempurl%
\url{https://doi.org/10.1016/j.jii.2017.02.003}
\showDOI{\tempurl}


\bibitem[\protect\citeauthoryear{De~Guglielmo, Brienza, and
  Anastasi}{De~Guglielmo et~al\mbox{.}}{2016}]%
        {de_guglielmo_ieee_2016}
\bibfield{author}{\bibinfo{person}{Domenico De~Guglielmo},
  \bibinfo{person}{Simone Brienza}, {and} \bibinfo{person}{Giuseppe Anastasi}.}
  \bibinfo{year}{2016}\natexlab{}.
\newblock \showarticletitle{{IEEE} 802.15.4e: {A} Survey}.
\newblock \bibinfo{journal}{\emph{Computer Communications}}
  \bibinfo{volume}{88} (\bibinfo{date}{Aug.} \bibinfo{year}{2016}),
  \bibinfo{pages}{1--24}.
\newblock
\showISSN{0140-3664}
\urldef\tempurl%
\url{https://doi.org/10.1016/j.comcom.2016.05.004}
\showDOI{\tempurl}


\bibitem[\protect\citeauthoryear{Domingo-Prieto, Chang, Vilajosana, and
  Watteyne}{Domingo-Prieto et~al\mbox{.}}{2016}]%
        {domingo-prieto_distributed_2016}
\bibfield{author}{\bibinfo{person}{Marc Domingo-Prieto},
  \bibinfo{person}{Tengfei Chang}, \bibinfo{person}{Xavier Vilajosana}, {and}
  \bibinfo{person}{Thomas Watteyne}.} \bibinfo{year}{2016}\natexlab{}.
\newblock \showarticletitle{Distributed {PID}-{Based} {Scheduling} for 6TiSCH
  {Networks}}.
\newblock \bibinfo{journal}{\emph{IEEE Communications Letters}}
  \bibinfo{volume}{20}, \bibinfo{number}{5} (\bibinfo{date}{May}
  \bibinfo{year}{2016}), \bibinfo{pages}{1006--1009}.
\newblock
\showISSN{1089-7798}
\urldef\tempurl%
\url{https://doi.org/10.1109/LCOMM.2016.2546880}
\showDOI{\tempurl}


\bibitem[\protect\citeauthoryear{Dujovne, Grieco, Palattella, and
  Accettura}{Dujovne et~al\mbox{.}}{2018}]%
        {sfx}
\bibfield{author}{\bibinfo{person}{Diego Dujovne}, \bibinfo{person}{Luigi
  Grieco}, \bibinfo{person}{Maria Palattella}, {and} \bibinfo{person}{Nicola
  Accettura}.} \bibinfo{year}{2018}\natexlab{}.
\newblock \bibinfo{title}{6TiSCH Experimental Scheduling Function (SFX)}.
\newblock
\newblock
\urldef\tempurl%
\url{http://www.ietf.org/internet-drafts/draft-ietf-6tisch-6top-sfx-01.txt}
\showURL{%
\tempurl}


\bibitem[\protect\citeauthoryear{Dunkels, Gr\"onvall, and Voigt}{Dunkels
  et~al\mbox{.}}{2004}]%
        {dunkels_contiki_2004}
\bibfield{author}{\bibinfo{person}{Adam Dunkels}, \bibinfo{person}{Bj\"orn
  Gr\"onvall}, {and} \bibinfo{person}{Thiemo Voigt}.}
  \bibinfo{year}{2004}\natexlab{}.
\newblock \showarticletitle{Contiki - {A} {Lightweight} and {Flexible}
  {Operating} {System} for {Tiny} {Networked} {Sensors}}. In
  \bibinfo{booktitle}{\emph{Proceedings of the 29th International Conference on
  Local Computer Networks}}. \bibinfo{publisher}{IEEE Computer Society},
  \bibinfo{address}{Los Alamitos, CA, USA}, \bibinfo{pages}{455--462}.
\newblock
\urldef\tempurl%
\url{https://doi.org/10.1109/LCN.2004.38}
\showDOI{\tempurl}


\bibitem[\protect\citeauthoryear{Duquennoy, Al~Nahas, Landsiedel, and
  Watteyne}{Duquennoy et~al\mbox{.}}{2015}]%
        {orchestra}
\bibfield{author}{\bibinfo{person}{Simon Duquennoy}, \bibinfo{person}{Beshr
  Al~Nahas}, \bibinfo{person}{Olaf Landsiedel}, {and} \bibinfo{person}{Thomas
  Watteyne}.} \bibinfo{year}{2015}\natexlab{}.
\newblock \showarticletitle{Orchestra: {Robust} {Mesh} {Networks} {Through}
  {Autonomously} {Scheduled} {TSCH}}. In \bibinfo{booktitle}{\emph{Proceedings
  of the 13th {ACM} {Conference} on {Embedded} {Networked} {Sensor}
  {Systems}}}. \bibinfo{pages}{337--350}.
\newblock
\showISBNx{978-1-4503-3631-4}
\urldef\tempurl%
\url{https://doi.org/10.1145/2809695.2809714}
\showDOI{\tempurl}


\bibitem[\protect\citeauthoryear{Gonga, Landsiedel, Soldati, and
  Johansson}{Gonga et~al\mbox{.}}{2012}]%
        {gonga_revisiting_2012}
\bibfield{author}{\bibinfo{person}{A. Gonga}, \bibinfo{person}{O. Landsiedel},
  \bibinfo{person}{P. Soldati}, {and} \bibinfo{person}{M. Johansson}.}
  \bibinfo{year}{2012}\natexlab{}.
\newblock \showarticletitle{Revisiting {Multi}-channel {Communication} to
  {Mitigate} {Interference} and {Link} {Dynamics} in {Wireless} {Sensor}
  {Networks}}. In \bibinfo{booktitle}{\emph{Proceedings of the 8th
  {International} {Conference} on {Distributed} {Computing} in {Sensor}
  {Systems}}}. \bibinfo{publisher}{IEEE}, \bibinfo{pages}{186--193}.
\newblock
\urldef\tempurl%
\url{https://doi.org/10.1109/DCOSS.2012.15}
\showDOI{\tempurl}


\bibitem[\protect\citeauthoryear{Hwang and Nam}{Hwang and Nam}{2014}]%
        {hwang_analysis_2014}
\bibfield{author}{\bibinfo{person}{Kwang-il Hwang} {and}
  \bibinfo{person}{Sung-wook Nam}.} \bibinfo{year}{2014}\natexlab{}.
\newblock \showarticletitle{{Analysis} and {Enhancement} of {IEEE} 802.15.4e
  {DSME} {Beacon} {Scheduling} {Model}}.
\newblock \bibinfo{journal}{\emph{Journal of Applied Mathematics}}
  \bibinfo{volume}{2014} (\bibinfo{date}{May} \bibinfo{year}{2014}).
\newblock
\showISSN{1110-757X}
\urldef\tempurl%
\url{https://doi.org/10.1155/2014/934610}
\showDOI{\tempurl}


\bibitem[\protect\citeauthoryear{Hwang, Wang, and Wang}{Hwang
  et~al\mbox{.}}{2017}]%
        {hwang_distributed_2017}
\bibfield{author}{\bibinfo{person}{Ren-Hung Hwang},
  \bibinfo{person}{Chih-Chiang Wang}, {and} \bibinfo{person}{Wu-Bin Wang}.}
  \bibinfo{year}{2017}\natexlab{}.
\newblock \showarticletitle{A {Distributed} {Scheduling} {Algorithm} for {IEEE}
  802.15.4e {Wireless} {Sensor} {Networks}}.
\newblock \bibinfo{journal}{\emph{Computer Standards \& Interfaces}}
  \bibinfo{volume}{52} (\bibinfo{date}{May} \bibinfo{year}{2017}),
  \bibinfo{pages}{63--70}.
\newblock
\showISSN{0920-5489}
\urldef\tempurl%
\url{https://doi.org/10.1016/j.csi.2017.01.003}
\showDOI{\tempurl}


\bibitem[\protect\citeauthoryear{Jeong and Lee}{Jeong and Lee}{2012}]%
        {jeong_performance_2012}
\bibfield{author}{\bibinfo{person}{Wun-Cheol Jeong} {and}
  \bibinfo{person}{Junhee Lee}.} \bibinfo{year}{2012}\natexlab{}.
\newblock \showarticletitle{Performance Evaluation of {IEEE} 802.15.4e {DSME}
  {MAC} Protocol for Wireless Sensor Networks}. In
  \bibinfo{booktitle}{\emph{Proceedings of the 1st {IEEE} {Workshop} on
  {Enabling} {Technologies} for {Smartphone} and {Internet} of {Things}
  ({ETSIoT} 2012)}}. \bibinfo{pages}{7--12}.
\newblock
\urldef\tempurl%
\url{https://doi.org/10.1109/ETSIoT.2012.6311258}
\showDOI{\tempurl}


\bibitem[\protect\citeauthoryear{Jeschke, Brecher, Song, and Rawat}{Jeschke
  et~al\mbox{.}}{2017}]%
        {jeschke_industrial_2017}
\bibfield{editor}{\bibinfo{person}{Sabina Jeschke}, \bibinfo{person}{Christian
  Brecher}, \bibinfo{person}{Houbing Song}, {and} \bibinfo{person}{Danda~B.
  Rawat}} (Eds.). \bibinfo{year}{2017}\natexlab{}.
\newblock \bibinfo{booktitle}{\emph{Industrial {Internet} of {Things}}}.
\newblock \bibinfo{publisher}{Springer}, \bibinfo{address}{Cham}.
\newblock
\showISBNx{978-3-319-42558-0}
\urldef\tempurl%
\url{https://doi.org/10.1007/978-3-319-42559-7}
\showDOI{\tempurl}


\bibitem[\protect\citeauthoryear{Juc, Alphand, Guizzetti, Favre, and Duda}{Juc
  et~al\mbox{.}}{2016}]%
        {juc_energy_2016}
\bibfield{author}{\bibinfo{person}{I. Juc}, \bibinfo{person}{O. Alphand},
  \bibinfo{person}{R. Guizzetti}, \bibinfo{person}{M. Favre}, {and}
  \bibinfo{person}{A. Duda}.} \bibinfo{year}{2016}\natexlab{}.
\newblock \showarticletitle{Energy Consumption and Performance of {IEEE}
  802.15.4e {TSCH} and {DSME}}. In \bibinfo{booktitle}{\emph{Proceedings of the
  2016 {IEEE} {Wireless} {Communications} and {Networking} {Conference}}}.
  \bibinfo{pages}{1--7}.
\newblock
\urldef\tempurl%
\url{https://doi.org/10.1109/WCNC.2016.7565006}
\showDOI{\tempurl}


\bibitem[\protect\citeauthoryear{Karp and Kung}{Karp and Kung}{2000}]%
        {karp_gpsr:_2000}
\bibfield{author}{\bibinfo{person}{Brad Karp} {and} \bibinfo{person}{H.~T.
  Kung}.} \bibinfo{year}{2000}\natexlab{}.
\newblock \showarticletitle{{GPSR}: {Greedy} {Perimeter} {Stateless} {Routing}
  for {Wireless} {Networks}}. In \bibinfo{booktitle}{\emph{Proceedings of the
  6th {Annual} {International} {Conference} on {Mobile} {Computing} and
  {Networking}}} \emph{(\bibinfo{series}{{MobiCom} '00})}.
  \bibinfo{publisher}{ACM}, \bibinfo{pages}{243--254}.
\newblock
\showISBNx{978-1-58113-197-0}
\urldef\tempurl%
\url{https://doi.org/10.1145/345910.345953}
\showDOI{\tempurl}


\bibitem[\protect\citeauthoryear{Kauer}{Kauer}{2018}]%
        {iotlab_consumption}
\bibfield{author}{\bibinfo{person}{Florian Kauer}.}
  \bibinfo{year}{2018}\natexlab{}.
\newblock \bibinfo{title}{{Per-Node Energy Consumption Measurements of the
  FIT/IoT-LAB}}.
\newblock
\newblock
\urldef\tempurl%
\url{https://doi.org/10.5281/zenodo.1288644}
\showDOI{\tempurl}


\bibitem[\protect\citeauthoryear{Kauer, K\"ostler, L\"ubkert, and Turau}{Kauer
  et~al\mbox{.}}{2016}]%
        {formaldsme}
\bibfield{author}{\bibinfo{person}{Florian Kauer}, \bibinfo{person}{Maximilian
  K\"ostler}, \bibinfo{person}{Tobias L\"ubkert}, {and} \bibinfo{person}{Volker
  Turau}.} \bibinfo{year}{2016}\natexlab{}.
\newblock \showarticletitle{{Formal Analysis and Verification of the IEEE
  802.15.4 DSME Slot Allocation}}. In \bibinfo{booktitle}{\emph{Proceedings of
  the 19th ACM International Conference on Modeling, Analysis and Simulation of
  Wireless and Mobile Systems (MSWIM 2016)}}.
\newblock
\urldef\tempurl%
\url{https://doi.org/10.1145/2988287.2989148}
\showDOI{\tempurl}


\bibitem[\protect\citeauthoryear{Kauer and Turau}{Kauer and Turau}{2018a}]%
        {modeltdma}
\bibfield{author}{\bibinfo{person}{Florian Kauer} {and} \bibinfo{person}{Volker
  Turau}.} \bibinfo{year}{2018}\natexlab{a}.
\newblock \showarticletitle{{An Analytical Model for Wireless Mesh Networks
  with Collision-Free TDMA and Finite Queues}}.
\newblock \bibinfo{journal}{\emph{EURASIP Journal on Wireless Communications
  and Networking}}  \bibinfo{volume}{2018} (\bibinfo{date}{June}
  \bibinfo{year}{2018}), \bibinfo{pages}{149}.
\newblock
\showISSN{1687-1499}
\urldef\tempurl%
\url{https://doi.org/10.1186/s13638-018-1146-x}
\showDOI{\tempurl}


\bibitem[\protect\citeauthoryear{Kauer and Turau}{Kauer and Turau}{2018b}]%
        {kauer_constructing_2018}
\bibfield{author}{\bibinfo{person}{Florian Kauer} {and} \bibinfo{person}{Volker
  Turau}.} \bibinfo{year}{2018}\natexlab{b}.
\newblock \showarticletitle{Constructing {Customized} {Multi}-{Hop}
  {Topologies} in {Dense} {Wireless} {Network} {Testbeds}}.
\newblock
\urldef\tempurl%
\url{http://arxiv.org/abs/1805.06661}
\showURL{%
\tempurl}
\newblock
\shownote{Accepted for the 17th International Conference on Ad Hoc Networks and
  Wireless (AdHoc-Now 2018). Preprint available.}


\bibitem[\protect\citeauthoryear{Kull}{Kull}{2015}]%
        {kull_mass_2015}
\bibfield{author}{\bibinfo{person}{Hans Kull}.}
  \bibinfo{year}{2015}\natexlab{}.
\newblock \bibinfo{booktitle}{\emph{Mass {Customization}}}.
\newblock \bibinfo{publisher}{Apress}, \bibinfo{address}{Berkeley, CA}.
\newblock
\showISBNx{978-1-4842-1008-6}
\urldef\tempurl%
\url{https://doi.org/10.1007/978-1-4842-1007-9}
\showDOI{\tempurl}


\bibitem[\protect\citeauthoryear{Lee and Jeong}{Lee and Jeong}{2012}]%
        {lee_performance_2012}
\bibfield{author}{\bibinfo{person}{J. Lee} {and} \bibinfo{person}{W.~C.
  Jeong}.} \bibinfo{year}{2012}\natexlab{}.
\newblock \showarticletitle{Performance Analysis of {IEEE} 802.15.4e {DSME}
  {MAC} Protocol under {WLAN} Interference}. In
  \bibinfo{booktitle}{\emph{Proceedings of the 3rd {International} {Conference}
  on {ICT} {Convergence} ({ICTC} 2012)}}. \bibinfo{pages}{741--746}.
\newblock
\urldef\tempurl%
\url{https://doi.org/10.1109/ICTC.2012.6387133}
\showDOI{\tempurl}


\bibitem[\protect\citeauthoryear{Lee, Exarchakos, and Liotta}{Lee
  et~al\mbox{.}}{2017}]%
        {lee_distributed_2017}
\bibfield{author}{\bibinfo{person}{T.~van~der Lee}, \bibinfo{person}{G.
  Exarchakos}, {and} \bibinfo{person}{A. Liotta}.}
  \bibinfo{year}{2017}\natexlab{}.
\newblock \showarticletitle{Distributed {TSCH} scheduling: {A} Comparative
  Analysis}. In \bibinfo{booktitle}{\emph{Proceedings of the 2017 {IEEE}
  {International} {Conference} on {Systems}, {Man}, and {Cybernetics} (IEEE
  {SMC} 2017)}}. \bibinfo{pages}{3517--3522}.
\newblock
\urldef\tempurl%
\url{https://doi.org/10.1109/SMC.2017.8123176}
\showDOI{\tempurl}


\bibitem[\protect\citeauthoryear{Lee and Chung}{Lee and Chung}{2016}]%
        {lee_efficient_2016}
\bibfield{author}{\bibinfo{person}{Yun-Sung Lee} {and}
  \bibinfo{person}{Sang-Hwa Chung}.} \bibinfo{year}{2016}\natexlab{}.
\newblock \showarticletitle{An {Efficient} {Distributed} {Scheduling}
  {Algorithm} for {Mobility} {Support} in {IEEE} 802.15.4e {DSME}-{Based}
  {Industrial} {Wireless} {Sensor} {Networks}}.
\newblock \bibinfo{journal}{\emph{International Journal of Distributed Sensor
  Networks}}  \bibinfo{volume}{2016} (\bibinfo{date}{Feb.}
  \bibinfo{year}{2016}).
\newblock
\showISSN{1550-1329}
\urldef\tempurl%
\url{https://doi.org/10.1155/2016/9837625}
\showDOI{\tempurl}


\bibitem[\protect\citeauthoryear{Liu, Li, Su, Fan, and Wang}{Liu
  et~al\mbox{.}}{2013}]%
        {liu_enhanced_2013}
\bibfield{author}{\bibinfo{person}{Xuecheng Liu}, \bibinfo{person}{Xiaoyun Li},
  \bibinfo{person}{Shijuan Su}, \bibinfo{person}{Zhenke Fan}, {and}
  \bibinfo{person}{Gang Wang}.} \bibinfo{year}{2013}\natexlab{}.
\newblock \showarticletitle{Enhanced {Fast} {Association} for 802.15.4e-2012
  {DSME} {MAC} {Protocol}}. In \bibinfo{booktitle}{\emph{Proceedings of the 2nd
  International Conference on Computer Science and Electronics Engineering}}.
  \bibinfo{publisher}{Atlantis Press}.
\newblock
\showISBNx{978-90-78677-61-1}
\urldef\tempurl%
\url{https://doi.org/10.2991/iccsee.2013.248}
\showDOI{\tempurl}


\bibitem[\protect\citeauthoryear{L\"ubkert}{L\"ubkert}{2015}]%
        {lubkert_joint_2015}
\bibfield{author}{\bibinfo{person}{Tobias L\"ubkert}.}
  \bibinfo{year}{2015}\natexlab{}.
\newblock \emph{\bibinfo{title}{Joint Analysis of {TDMA} and Geographic Routing
  for a Large Scale Wireless Mesh Network}}.
\newblock Master's Thesis. \bibinfo{school}{Technische Universit\"at Hamburg}.
\newblock
\urldef\tempurl%
\url{https://doi.org/10.15480/882.1664}
\showDOI{\tempurl}


\bibitem[\protect\citeauthoryear{Meier and Turau}{Meier and Turau}{2015a}]%
        {meiermodel}
\bibfield{author}{\bibinfo{person}{Florian Meier} {and} \bibinfo{person}{Volker
  Turau}.} \bibinfo{year}{2015}\natexlab{a}.
\newblock \showarticletitle{{An Analytical Model for Fast and Verifiable
  Assessment of Large Scale Wireless Mesh Networks}}. In
  \bibinfo{booktitle}{\emph{{Proceedings of the 11th International Conference
  on the Design of Reliable Communication Networks (DRCN)}}}.
  \bibinfo{publisher}{IEEE}.
\newblock
\urldef\tempurl%
\url{https://doi.org/10.1109/DRCN.2015.7149011}
\showDOI{\tempurl}


\bibitem[\protect\citeauthoryear{Meier and Turau}{Meier and Turau}{2015b}]%
        {modelarxiv}
\bibfield{author}{\bibinfo{person}{Florian Meier} {and} \bibinfo{person}{Volker
  Turau}.} \bibinfo{year}{2015}\natexlab{b}.
\newblock \showarticletitle{{Analytical Model for IEEE 802.15.4 Multi-Hop
  Networks with Improved Handling of Acknowledgements and Retransmissions}}.
\newblock  (\bibinfo{year}{2015}).
\newblock
\showeprint{1501.07594}
\urldef\tempurl%
\url{http://arxiv.org/abs/1501.07594}
\showURL{%
\tempurl}


\bibitem[\protect\citeauthoryear{Nam and Hwang}{Nam and Hwang}{2014}]%
        {nam_enhanced_2014}
\bibfield{author}{\bibinfo{person}{Sung-wook Nam} {and}
  \bibinfo{person}{Kwang-il Hwang}.} \bibinfo{year}{2014}\natexlab{}.
\newblock \showarticletitle{Enhanced {Beacon} {Scheduling} of {IEEE}802.15.4e
  {DSME}}.
\newblock In \bibinfo{booktitle}{\emph{Frontier and {Innovation} in {Future}
  {Computing} and {Communications}}}. \bibinfo{publisher}{Springer, Dordrecht},
  \bibinfo{pages}{495--503}.
\newblock
\urldef\tempurl%
\url{https://doi.org/10.1007/978-94-017-8798-7_60}
\showDOI{\tempurl}


\bibitem[\protect\citeauthoryear{Ojo, Giordano, Portaluri, Adami, and
  Pagano}{Ojo et~al\mbox{.}}{2017}]%
        {ojo_energy_2017}
\bibfield{author}{\bibinfo{person}{M. Ojo}, \bibinfo{person}{S. Giordano},
  \bibinfo{person}{G. Portaluri}, \bibinfo{person}{D. Adami}, {and}
  \bibinfo{person}{M. Pagano}.} \bibinfo{year}{2017}\natexlab{}.
\newblock \showarticletitle{An Energy Efficient Centralized Scheduling Scheme
  in {TSCH} Networks}. In \bibinfo{booktitle}{\emph{Proceedings of the 2017
  {IEEE} {International} {Conference} on {Communications} {Workshops} ({ICC}
  {Workshops})}}. \bibinfo{pages}{570--575}.
\newblock
\urldef\tempurl%
\url{https://doi.org/10.1109/ICCW.2017.7962719}
\showDOI{\tempurl}


\bibitem[\protect\citeauthoryear{OMNeT++}{OMNeT++}{2018}]%
        {INET}
\bibfield{author}{\bibinfo{person}{OMNeT++}.} \bibinfo{year}{2018}\natexlab{}.
\newblock \bibinfo{title}{{INET Framework for OMNeT++}}.
\newblock \bibinfo{howpublished}{\url{https://inet.omnetpp.org/}}.
\newblock


\bibitem[\protect\citeauthoryear{Palattella, Accettura, Grieco, Boggia, Dohler,
  and Engel}{Palattella et~al\mbox{.}}{2013}]%
        {palattella_optimal_2013}
\bibfield{author}{\bibinfo{person}{M.~R. Palattella}, \bibinfo{person}{N.
  Accettura}, \bibinfo{person}{L.~A. Grieco}, \bibinfo{person}{G. Boggia},
  \bibinfo{person}{M. Dohler}, {and} \bibinfo{person}{T. Engel}.}
  \bibinfo{year}{2013}\natexlab{}.
\newblock \showarticletitle{On {Optimal} {Scheduling} in {Duty}-{Cycled}
  {Industrial} {IoT} {Applications} {Using} {IEEE}802.15.4e {TSCH}}.
\newblock \bibinfo{journal}{\emph{IEEE Sensors Journal}} \bibinfo{volume}{13},
  \bibinfo{number}{10} (\bibinfo{date}{Oct.} \bibinfo{year}{2013}),
  \bibinfo{pages}{3655--3666}.
\newblock
\showISSN{1530-437X}
\urldef\tempurl%
\url{https://doi.org/10.1109/JSEN.2013.2266417}
\showDOI{\tempurl}


\bibitem[\protect\citeauthoryear{Palattella, Watteyne, Wang, Muraoka,
  Accettura, Dujovne, Grieco, and Engel}{Palattella et~al\mbox{.}}{2016}]%
        {palattella_--fly_2016}
\bibfield{author}{\bibinfo{person}{M.~R. Palattella}, \bibinfo{person}{T.
  Watteyne}, \bibinfo{person}{Q. Wang}, \bibinfo{person}{K. Muraoka},
  \bibinfo{person}{N. Accettura}, \bibinfo{person}{D. Dujovne},
  \bibinfo{person}{L.~A. Grieco}, {and} \bibinfo{person}{T. Engel}.}
  \bibinfo{year}{2016}\natexlab{}.
\newblock \showarticletitle{On-the-{Fly} {Bandwidth} {Reservation} for 6TiSCH
  {Wireless} {Industrial} {Networks}}.
\newblock \bibinfo{journal}{\emph{IEEE Sensors Journal}} \bibinfo{volume}{16},
  \bibinfo{number}{2} (\bibinfo{date}{Jan.} \bibinfo{year}{2016}),
  \bibinfo{pages}{550--560}.
\newblock
\showISSN{1530-437X}
\urldef\tempurl%
\url{https://doi.org/10.1109/JSEN.2015.2480886}
\showDOI{\tempurl}


\bibitem[\protect\citeauthoryear{Pfahl, Randt, Meier, Zaschke, Geurts, and
  Buselmeier}{Pfahl et~al\mbox{.}}{2014}]%
        {pfahl_holistic_2014}
\bibfield{author}{\bibinfo{person}{Andreas Pfahl}, \bibinfo{person}{Michael
  Randt}, \bibinfo{person}{Florian Meier}, \bibinfo{person}{Martin Zaschke},
  \bibinfo{person}{C.~P.~W. Geurts}, {and} \bibinfo{person}{Michael
  Buselmeier}.} \bibinfo{year}{2014}\natexlab{}.
\newblock \showarticletitle{{A Holistic Approach for Low Cost Heliostat
  Fields}}. In \bibinfo{booktitle}{\emph{Proceedings of the 20th
  {International} {Conference} on {Concentrated} {Solar} {Power} and {Chemical}
  {Energy} {Technologies} {(SolarPACES 2014)}}}. \bibinfo{address}{Peking,
  China}.
\newblock
\urldef\tempurl%
\url{https://doi.org/10.1016/j.egypro.2015.03.021}
\showDOI{\tempurl}


\bibitem[\protect\citeauthoryear{Polastre, Hill, and Culler}{Polastre
  et~al\mbox{.}}{2004}]%
        {polastre_versatile_2004}
\bibfield{author}{\bibinfo{person}{Joseph Polastre}, \bibinfo{person}{Jason
  Hill}, {and} \bibinfo{person}{David Culler}.}
  \bibinfo{year}{2004}\natexlab{}.
\newblock \showarticletitle{Versatile {Low} {Power} {Media} {Access} for
  {Wireless} {Sensor} {Networks}}. In \bibinfo{booktitle}{\emph{Proceedings of
  the 2nd {International} {Conference} on {Embedded} {Networked} {Sensor}
  {Systems}}} \emph{(\bibinfo{series}{{SenSys} '04})}.
  \bibinfo{publisher}{ACM}, \bibinfo{pages}{95--107}.
\newblock
\showISBNx{978-1-58113-879-5}
\urldef\tempurl%
\url{https://doi.org/10.1145/1031495.1031508}
\showDOI{\tempurl}


\bibitem[\protect\citeauthoryear{P\"ottner, Seidel, Brown, Roedig, and
  Wolf}{P\"ottner et~al\mbox{.}}{2014}]%
        {pottner_constructing_2014}
\bibfield{author}{\bibinfo{person}{Wolf-Bastian P\"ottner},
  \bibinfo{person}{Hans Seidel}, \bibinfo{person}{James Brown},
  \bibinfo{person}{Utz Roedig}, {and} \bibinfo{person}{Lars Wolf}.}
  \bibinfo{year}{2014}\natexlab{}.
\newblock \showarticletitle{Constructing {Schedules} for {Time}-{Critical}
  {Data} {Delivery} in {Wireless} {Sensor} {Networks}}.
\newblock \bibinfo{journal}{\emph{ACM Transactions on Sensor Networks}}
  \bibinfo{volume}{10}, \bibinfo{number}{3} (\bibinfo{date}{May}
  \bibinfo{year}{2014}), \bibinfo{pages}{44:1--44:31}.
\newblock
\showISSN{1550-4859}
\urldef\tempurl%
\url{https://doi.org/10.1145/2494528}
\showDOI{\tempurl}


\bibitem[\protect\citeauthoryear{Sahoo, Pattanaik, and Wu}{Sahoo
  et~al\mbox{.}}{2017}]%
        {sahoo_novel_2017}
\bibfield{author}{\bibinfo{person}{Prasan Sahoo}, \bibinfo{person}{Sudhir
  Pattanaik}, {and} \bibinfo{person}{Shih-Lin Wu}.}
  \bibinfo{year}{2017}\natexlab{}.
\newblock \showarticletitle{A {Novel} {IEEE} 802.15.4e {DSME} {MAC} for
  {Wireless} {Sensor} {Networks}}.
\newblock \bibinfo{journal}{\emph{Sensors}} \bibinfo{volume}{17},
  \bibinfo{number}{1} (\bibinfo{date}{Jan.} \bibinfo{year}{2017}),
  \bibinfo{pages}{168}.
\newblock
\showISSN{1424-8220}
\urldef\tempurl%
\url{https://doi.org/10.3390/s17010168}
\showDOI{\tempurl}


\bibitem[\protect\citeauthoryear{{Stefan Untersch\"utz, Andreas Weigel and
  Volker Turau}}{{Stefan Untersch\"utz, Andreas Weigel and Volker
  Turau}}{2012}]%
        {unterschutz2012cross}
\bibfield{author}{\bibinfo{person}{{Stefan Untersch\"utz, Andreas Weigel and
  Volker Turau}}.} \bibinfo{year}{2012}\natexlab{}.
\newblock \showarticletitle{Cross-Platform Protocol Development Based on
  OMNeT++}. In \bibinfo{booktitle}{\emph{Proceedings of the 5th International
  ICST Conference on Simulation Tools and Techniques (SIMUTOOLS 2012)}}.
  \bibinfo{pages}{278--282}.
\newblock
\urldef\tempurl%
\url{https://dl.acm.org/citation.cfm?id=2263063}
\showURL{%
\tempurl}


\bibitem[\protect\citeauthoryear{Theoleyre and Papadopoulos}{Theoleyre and
  Papadopoulos}{2016}]%
        {theoleyre_experimental_2016}
\bibfield{author}{\bibinfo{person}{Fabrice Theoleyre} {and}
  \bibinfo{person}{Georgios~Z. Papadopoulos}.} \bibinfo{year}{2016}\natexlab{}.
\newblock \showarticletitle{Experimental {Validation} of a {Distributed}
  {Self}-{Configured} 6TiSCH with {Traffic} {Isolation} in {Low} {Power}
  {Lossy} {Networks}}. In \bibinfo{booktitle}{\emph{Proceedings of the 19th
  {ACM} {International} {Conference} on {Modeling}, {Analysis} and {Simulation}
  of {Wireless} and {Mobile} {Systems} (MSWIM 2016)}}.
  \bibinfo{pages}{102--110}.
\newblock
\urldef\tempurl%
\url{https://doi.org/10.1145/2988287.2989133}
\showDOI{\tempurl}


\bibitem[\protect\citeauthoryear{Thubert}{Thubert}{2017}]%
        {ietf-6tisch-architecture-13}
\bibfield{author}{\bibinfo{person}{Pascal Thubert}.}
  \bibinfo{year}{2017}\natexlab{}.
\newblock \bibinfo{booktitle}{\emph{{An Architecture for IPv6 over the TSCH
  mode of IEEE 802.15.4}}}.
\newblock \bibinfo{type}{Internet-Draft} draft-ietf-6tisch-architecture-13.
  \bibinfo{institution}{Internet Engineering Task Force}.
\newblock
\urldef\tempurl%
\url{https://datatracker.ietf.org/doc/html/draft-ietf-6tisch-architecture-13}
\showURL{%
\tempurl}
\newblock
\shownote{Work in Progress.}


\bibitem[\protect\citeauthoryear{Tinka, Watteyne, Pister, and Bayen}{Tinka
  et~al\mbox{.}}{2011}]%
        {tinka_decentralized_2011}
\bibfield{author}{\bibinfo{person}{Andrew Tinka}, \bibinfo{person}{Thomas
  Watteyne}, \bibinfo{person}{Kristofer S.~J. Pister}, {and}
  \bibinfo{person}{Alexandre~M. Bayen}.} \bibinfo{year}{2011}\natexlab{}.
\newblock \showarticletitle{A Decentralized Scheduling Algorithm for Time
  Synchronized Channel Hopping}.
\newblock \bibinfo{journal}{\emph{EAI Endorsed Transactions on Mobile
  Communications and Applications}} \bibinfo{volume}{11}, \bibinfo{number}{1}
  (\bibinfo{date}{Sept.} \bibinfo{year}{2011}).
\newblock
\showISSN{2032-9504}
\urldef\tempurl%
\url{https://doi.org/10.4108/icst.trans.mca.2011.e5}
\showDOI{\tempurl}


\bibitem[\protect\citeauthoryear{Vallati, Brienza, Palmieri, and
  Anastasi}{Vallati et~al\mbox{.}}{2017}]%
        {vallati_improving_2017}
\bibfield{author}{\bibinfo{person}{Carlo Vallati}, \bibinfo{person}{Simone
  Brienza}, \bibinfo{person}{Maurizio Palmieri}, {and}
  \bibinfo{person}{Giuseppe Anastasi}.} \bibinfo{year}{2017}\natexlab{}.
\newblock \showarticletitle{Improving Network Formation in {IEEE} 802.15.4e
  {DSME}}.
\newblock \bibinfo{journal}{\emph{Computer Communications}}
  \bibinfo{volume}{114} (\bibinfo{date}{Dec.} \bibinfo{year}{2017}),
  \bibinfo{pages}{1--9}.
\newblock
\showISSN{0140-3664}
\urldef\tempurl%
\url{https://doi.org/10.1016/j.comcom.2017.09.016}
\showDOI{\tempurl}


\bibitem[\protect\citeauthoryear{Varga}{Varga}{2001}]%
        {varga2001omnet++}
\bibfield{author}{\bibinfo{person}{Andr\'as Varga}.}
  \bibinfo{year}{2001}\natexlab{}.
\newblock \showarticletitle{The OMNeT++ Discrete Event Simulation System}. In
  \bibinfo{booktitle}{\emph{Proceedings of the European Simulation
  Multiconference (ESM 2001)}}.
\newblock
\urldef\tempurl%
\url{https://pdfs.semanticscholar.org/0586/f39a5280d49e62b49838c229dcb37d105994.pdf}
\showURL{%
\tempurl}


\bibitem[\protect\citeauthoryear{Wan, Eisenman, Campbell, and Crowcroft}{Wan
  et~al\mbox{.}}{2005}]%
        {wan_siphon:_2005}
\bibfield{author}{\bibinfo{person}{Chieh-Yih Wan}, \bibinfo{person}{Shane~B.
  Eisenman}, \bibinfo{person}{Andrew~T. Campbell}, {and} \bibinfo{person}{Jon
  Crowcroft}.} \bibinfo{year}{2005}\natexlab{}.
\newblock \showarticletitle{Siphon: {Overload} Traffic Management Using
  Multi-Radio Virtual Sinks in Sensor Networks}. In
  \bibinfo{booktitle}{\emph{Proceedings of the 3rd International Conference on
  Embedded Networked Sensor Systems (SenSys 2005)}}. \bibinfo{pages}{116--129}.
\newblock
\urldef\tempurl%
\url{https://doi.org/10.1145/1098918.1098931}
\showDOI{\tempurl}


\bibitem[\protect\citeauthoryear{Wang, Li, and Zhao}{Wang
  et~al\mbox{.}}{2009}]%
        {wang_analysis_2009}
\bibfield{author}{\bibinfo{person}{F. Wang}, \bibinfo{person}{D. Li}, {and}
  \bibinfo{person}{Y. Zhao}.} \bibinfo{year}{2009}\natexlab{}.
\newblock \showarticletitle{Analysis and {Compare} of {Slotted} and {Unslotted}
  {CSMA} in {IEEE} 802.15.4}. In \bibinfo{booktitle}{\emph{Proceedings of the
  5th {International} {Conference} on {Wireless} {Communications}, {Networking}
  and {Mobile} {Computing}}}. \bibinfo{pages}{1--5}.
\newblock
\urldef\tempurl%
\url{https://doi.org/10.1109/WICOM.2009.5303580}
\showDOI{\tempurl}


\bibitem[\protect\citeauthoryear{Watteyne, Vilajosana, Kerkez, Chraim, Weekly,
  Wang, Glaser, and Pister}{Watteyne et~al\mbox{.}}{2012}]%
        {watteyne_openwsn:_2012}
\bibfield{author}{\bibinfo{person}{Thomas Watteyne}, \bibinfo{person}{Xavier
  Vilajosana}, \bibinfo{person}{Branko Kerkez}, \bibinfo{person}{Fabien
  Chraim}, \bibinfo{person}{Kevin Weekly}, \bibinfo{person}{Qin Wang},
  \bibinfo{person}{Steven Glaser}, {and} \bibinfo{person}{Kris Pister}.}
  \bibinfo{year}{2012}\natexlab{}.
\newblock \showarticletitle{{OpenWSN}: A Standards-Based Low-Power Wireless
  Development Environment}.
\newblock \bibinfo{journal}{\emph{Transactions on Emerging Telecommunications
  Technologies}} \bibinfo{volume}{23}, \bibinfo{number}{5}
  (\bibinfo{date}{Aug.} \bibinfo{year}{2012}).
\newblock
\showISSN{2161-3915}
\urldef\tempurl%
\url{https://doi.org/10.1002/ett.2558}
\showDOI{\tempurl}


\bibitem[\protect\citeauthoryear{Weigel and Turau}{Weigel and Turau}{2015}]%
        {Weigel2015}
\bibfield{author}{\bibinfo{person}{Andreas Weigel} {and}
  \bibinfo{person}{Volker Turau}.} \bibinfo{year}{2015}\natexlab{}.
\newblock \showarticletitle{{Hardware-Assisted IEEE 802.15.4 Transmissions and
  Why to Avoid Them}}. In \bibinfo{booktitle}{\emph{{Proceedings of the 8th
  International Conference on Internet and Distributed Computing Systems (IDCS
  2015)}}}. \bibinfo{publisher}{Springer}, \bibinfo{address}{Cham},
  \bibinfo{pages}{223--234}.
\newblock
\showISBNx{978-3-319-23237-9}
\urldef\tempurl%
\url{https://doi.org/10.1007/978-3-319-23237-9_20}
\showDOI{\tempurl}


\end{thebibliography}

\label{lastpage} 
\end{document}